\documentclass[pra,twocolumn,floatfix,a4paper,superscriptaddress]{revtex4}
\usepackage{bm,color,graphicx,amsmath,txfonts}

\usepackage[colorlinks, citecolor=blue,linkcolor=blue]{hyperref}

\begin{document}
	
\title{Generated multi-transparency windows, Fano resonances, and slow/fast light in an opto-magnomechanical system through magnon Squeezing}
	
\author{M$'$bark Amghar}
\affiliation{LPTHE-Department of Physics, Faculty of sciences, Ibnou Zohr University, Agadir, Morocco}

\author{Mohamed Amazioug}\email{m.amazioug@uiz.ac.ma}
\affiliation{LPTHE-Department of Physics, Faculty of sciences, Ibnou Zohr University, Agadir, Morocco}

\begin{abstract}
		
This study presents a theoretical investigation of magnomechanically induced transparency, Fano resonance, and slow/fast light phenomena within a hybrid optomagnomechanical system. The system consists of a mechanical membrane within a cavity and a ferrimagnetic crystal with a magnon squeezing mode that couples directly to a microwave cavity mode through magnetic dipole interaction and indirectly to an optical cavity through the crystal's deformation displacement. The optical cavity is influenced by the mechanical displacement, which is driven by magnetostrictive forces through radiation pressure. The interactions among photon, phonons, magnons, and microwave within the cavity give rise to optomechanically induced transparency (OMIT), magnomechanically induced transparency (MMIT) and magnon-induced transparency (MIT). We analyze how the presence of magnon squeezing modifies the absorption and dispersion properties of the optical spectrum. Additionally, the emergence of Fano resonance via photon-phonon coupling in the presence of the magnon squeezing is examined. By tuning the microwave-magnon interactions and the magnon squeezing parameters, we enhance slow/light effects. Our results aim to contribute to the development of quantum information processing technologies. 

\end{abstract}
	
	\date{\today}
	
	\maketitle

	\section{INTRODUCTION}
	
The light-matter interaction has emerged as a substantial research area with promising applications in a variety of fields. These include electromagnetically induced transparency (EIT) \cite{1,2,3}, quantum entanglement \cite{4,5,6}, macroscopic quantum superposition states \cite{7}, photon and magnon blockades \cite{8,9,10}, optomechanically induced transparency (OMIT) \cite{11,12,13,14,15,16}, and squeezing \cite{17,18}. Over the past decade, this field has driven progress in developing robust quantum memory and information processing technologies. EIT, a quantum interference effect, is observed in three-level atomic systems \cite{19}. It suppresses atomic absorption to zero using an auxiliary laser field, primarily through interference effects or resonance in a dark state. OMIT, resulting from the interference between a weak probe field and an anti-Stokes scattering field \cite{20}, is a similar phenomenon, both experimentally \cite{15,16} and theoretically \cite{13} examined. The emergence of new optomechanical cavity systems has spurred theoretical research into multiple OMITs within atomic support-assisted and hybrid piezo-optomechanical systems \cite{21,22,23}, as well as in multiple resonator optomechanical frameworks \cite{24}. Furthermore, the complexity of light-matter interactions is exemplified by Fano resonance, originally found in atomic systems \cite{25}. This phenomenon is the consequence of quantum interference of various transition amplitudes, resulting in absorption profile minima \cite{27}. It has been investigated in a variety of physical systems, including optomechanical systems \cite{12,28}, coupled microresonators \cite{29} and photonic crystals \cite{30}. A hybrid-cavity magnomechanical system has been observed to exhibit asymmetric profiles similar to those of Fano in recent experiments \cite{31,32}. \\

The cavity magnomechanics system (CMM) \cite{32} has recently attracted more attention than cavity optomechanics due to its numerous advantages over conventional systems. Magnetic materials are known for their high spin density and extremely low damping rate \cite{33,34}, which makes them an ideal framework for the study of intense interactions between light and matter. Besides the advantages mentioned above, the magnon is highly capable of interacting with various quantum systems, like optical photons, phonons \cite{36}, microwaves \cite{37}, superconducting qubits \cite{38}, and whispering gallery modes (WGMs) \cite{39}. Because of these special characteristics, YIG-containing systems are employed to investigate a diverse array of coherent phenomena that are comparable to those of optomechanical systems, including magnomechanically induced transparency (MMIT) \cite{40}, magnomechanical cooling and entanglement \cite{41,42,43,44}, fast/slow light engineering \cite{45,46}, bistability and non-reciprocity \cite{47,48}, ground-state cooling of magnomechanical resonators \cite{49,50}, and so on.	\\

The innovative system of opto-magnomechanics is the result of the integration of cavity optomechanics and cavity magnomechanics, which is accomplished by linking the magnomechanical displacement to an optical cavity through radiation pressure \cite{51,52,53}. This hybrid system enables the optical detection of magnon populations \cite{51}, the manipulation of various magnonic quantum states in solid materials \cite{53}, and the generation of entanglement between optomagnonics and microwave optics \cite{52}. As a result, it shows significant potential for applications in quantum information processing and the advancement of quantum networks \cite{54}.\\

In recent years, opto-magnomechanical systems have successfully demonstrated magnomechanically induced transparency (MMIT), slow/fast light, and entanglement. For example, our most recent research investigated the slow/fast light effect, Fano resonance, and magnomechanically induced transparency in situations where an atomic ensemble is located within the hybrid cavity of an opto-magnomechanical system \cite{12}. In addition, Fan et al. leveraged nonlinear magnetostrictive and radiation-pressure interactions to investigate entanglement within a cavity opto-magnomechanical (OMM) configuration. A magnon mode in a ferrimagnetic crystal, such as yttrium-iron-garnet (YIG), is dispersively coupled to a phonon mode that is the result of magnetostrictively induced mechanical vibrations in the OMM system. Furthermore, this mechanical motion is further coupled to an optical cavity through radiation pressure, and the magnon mode interacts with a microwave cavity mode through the magnetic dipole interaction, resulting in the formation of the cavity-OMM system \cite{52}.\\

Building on the preceding studies, we will explore the MMIT phenomenon, Fano resonance, and the fast-slow light effect within an opto-magnomechanical system comprising a mechanical membrane and a magnon squeezing

mode within a ferrimagnetic crystal, such as yttrium iron garnet (YIG). This magnon mode is dispersively coupled to a phonon mode generated by magnetostrictive mechanical vibrations. The mechanical motion is further coupled to an optical cavity through radiation pressure. Additionally, the magnon mode interacts with a microwave cavity mode via magnetic dipole interaction, thereby constituting the cavity-OMM system. Our study focuses on the impact of different couplings and squeezing parameter on absorption and dispersion spectra. Additionally, we will examine the appearance of Fano resonance in the output field and identify the optimal system parameters for its observation. Moreover, we will discuss slow/fast light propagation, emphasizing that the group delay is influenced by the tunability of the microwave–magnon coupling and the squeezing parameter in the YIG sphere.\\

The paper is structured as follows: Section \ref{o} introduces the formula for the Hamiltonian and the relevant Quantum Langevin Equations for the system under study, along with deriving the output field. Section \ref{I} explores MMIT and analyzes how the different couplings strength affects the input spectrum. In Section \ref{II}, we identify the critical conditions necessary for the occurrence of double Fano resonances in our hybrid system. Section \ref{III} examines the group delay associated with slow and fast light propagation. We conclude with final remarks.

\begin{figure}[t]
	\centering
	\includegraphics[width=0.48\textwidth]{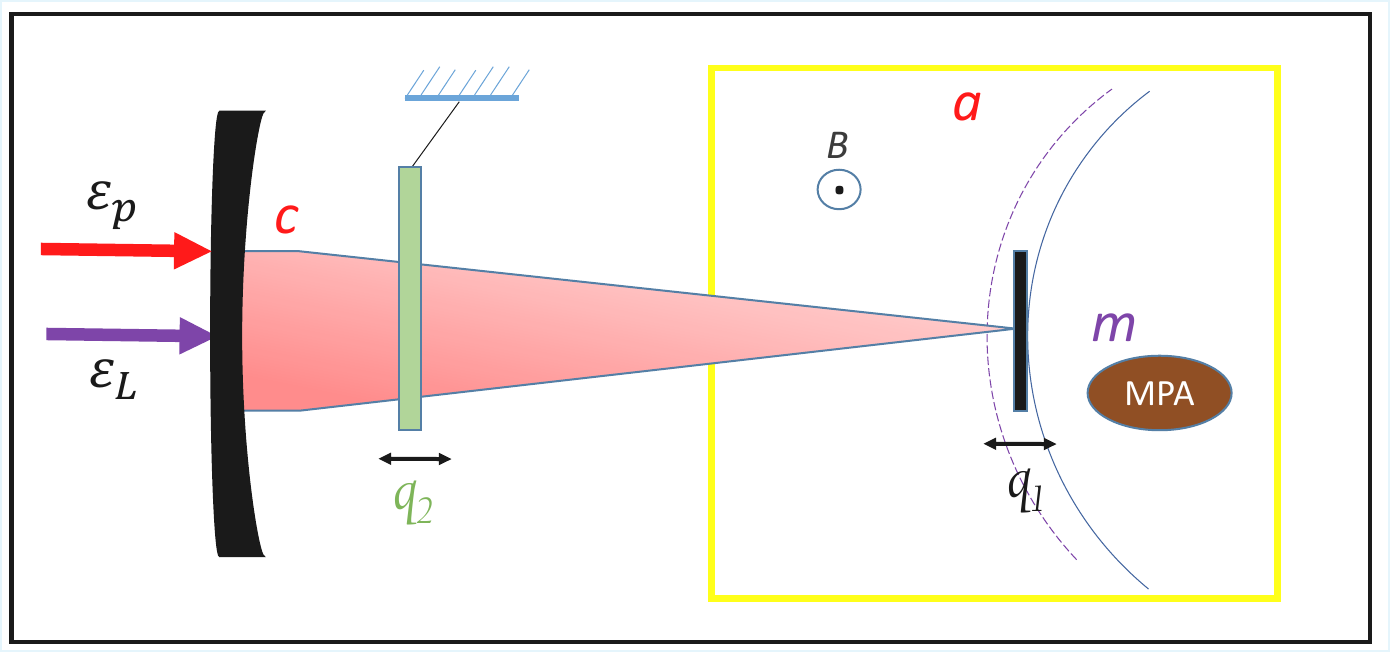}
	\caption{Diagram illustrating the opto-magnomechanical system, where an optical cavity mode $c$, driven by a probe field at frequency $\omega_p$, is coupled to a mechanical membrane $b_2$ and to a magnon squeezing mode $m$ in a YIG crystal via the mechanical vibration $b_1$ induced by the magnetostriction. In addition, a magnon mode is coupled to a microwave cavity mode $a$.}\label{1}
\end{figure}
\section{SYSTEM AND MODEL}\label{o}
A cavity-OMM system, as illustrated in Figure \eqref{1}, is the subject of our investigation. This system is comprised of an optical cavity mode, a mechanical vibration mode, a magnon mode in a YIG crystal, and a microwave cavity mode. The magnon mode is represented in the YIG crystal by the collective motion (spin wave) of a large number of spins. The crystal is activated by placing it in a uniform bias magnetic field and applying a microwave drive field with its magnetic component perpendicular to the bias field. The magnon mode couples to a microwave cavity field by positioning the crystal in close proximity to the cavity mode's maximal magnetic field through the magnetic dipole interaction \cite{II1,II2}. In a relatively large crystal, the magnetostriction interaction of the ferrimagnet results in a dispersive coupling between magnons and lower-frequency vibrational phonons \cite{32,36,44,47}. This magnomechanical displacement is further coupled to an optical cavity through radiation pressure \cite{53,II8}. The YIG crystal can be formed into either a sphere \cite{32,44,47} or a micro bridge structure \cite{53,II9}. For ease in fabricating an optical cavity, we choose the micro bridge structure. An optical cavity is created by attaching a small, highly-reflective mirror to the surface of the micro bridge \cite{51,II11}. This mirror must be compact and lightweight to prevent it from substantially altering the mechanical properties of the micro bridge, including its displacement. The bending displacement can be minimized by assuming that the displacement is uniform and perpendicular to the affixed surface. This enables the YIG bridge and mirror to oscillate at nearly the same frequency and adhere closely. Alternatively, the 'membrane-in-the-middle' method may be implemented for optomechanical dispersive coupling \cite{II12}. A smaller micro bridge may provide greater magnomechanical coupling. However, we avoid using thin YIG films where high-frequency phonon modes have strong linear coupling with the magnon mode. The Hamiltonian that describes this system is given as
\begin{equation}
	\begin{aligned}
		\mathcal{H} =\mathcal{ H}_{{free}}+\mathcal{H}_{{int}}+\mathcal{H}_{{drive}}+\mathcal{H}_{{MPA}},
	\end{aligned}
\end{equation}
where $(j=a,m,c)$
\begin{equation}
	\mathcal{H}_{{free}}=\sum_{j=c, m, a} \omega_j j^{\dagger} j+\frac{{\omega_b}_1}{2}\left(q_1^2+p_1^2\right)+\frac{{\omega_b}_2}{2}\left(q_2^2+p_2^2\right),
\end{equation}
\begin{equation}
	\mathcal{H}_{{int}}=g_a\left(a^{\dagger} m+a m^{\dagger}\right)+g_m m^{\dagger} m q_1-g_1 c^{\dagger} c q_1-g_2 c^{\dagger} c q_2,	
\end{equation}
\begin{equation}
\begin{aligned}
	\mathcal{H}_{{drive}}=&i \Omega\left(m^{\dagger} e^{-i \omega_L t}-\text { H.c }\right)+i \epsilon_L\left(c^{\dagger} e^{-i \omega_L t}-\text { H.c}\right)\\
	&+i \epsilon_p\left(c^{\dagger} e^{-i \omega_p t}-\text { H.c }\right),
\end{aligned}
\end{equation}
\begin{equation}
	\mathcal{H}_{{MPA}}=i\lambda\left(e^{i\theta} {m^\dagger}^2 e^{-2i\omega_Lt}-e^{-i\theta} m^2e^{2i\omega_Lt}\right).
\end{equation}
The free Hamiltonian is denoted by $\mathcal{H}_{{free}}$, with the first term describing the energy of the optical cavity mode, the magnon mode, and the microwave cavity mode, respectively. The resonant frequencies of the three modes are denoted by $\omega_j$ $(j = c, m, a)$. The magnon frequency is determined by the gyromagnetic ratio $\gamma_r$ and the external bias magnetic field $H$, where $\gamma_r/2\pi=28$ GHz/T \cite{44}. More specifically, $\omega_{m}=\gamma_r H$. $j(j^\dagger)$ are the annihilation (creation) operators that satisfy the standard commutation relation $[j,j^\dagger] = 1$, where $j = a,m,c$. The energy of two mechanical vibration modes, with frequencies $\omega_{b1}$ and $\omega_{b2}$, is represented by the second and third terms. Each vibration mode is modeled as a mechanical oscillator, with $[q_1, p_1] = i$ and $[q_2, p_2] = i$. The dimensionless position and momentum of these modes are $q_1$ and $p_1$, and $q_2$ and $p_2$, respectively. $\mathcal{H}_{{int}}$ denotes the energy of the interaction between different modes, with $g_a$ is the microwave-magnon coupling rate and $g_m$ is the magnomechanical coupling rate, which can be significantly enhanced by driving the magnon mode with a strong field. $g_1$ and $g_2$ represent the mechanical phonon-photon coupling rates. The drive Hamiltonian represent by $\mathcal{H}_{{drive}}$. An external laser field with an amplitude of $\epsilon_L=\sqrt{2 \kappa_{c} P_L / \hbar \omega_L}$ is used to drive the optical cavity where $\kappa_{c}$ is the decay rate of the optical cavity mode and $\omega_L$ $(P_L)$ is the frequency (power) of the external input laser field. Concurrently, the cavity mode is motivated by a feeble probe field with amplitude $\epsilon_p$ and frequency $\omega_p$. At frequency \( \omega_L \), the Rabi frequency \( \Omega = \sqrt{5}/4 \gamma_0 \sqrt{N_s }B_d \) \cite{44} denotes the coupling strength of the input laser drive field. Here, $ \gamma_0 $ is the gyromagnetic ratio, $ N_s $ is the total spin number of the ferrimagnet, and $B_d $ is the amplitude of the drive magnetic field. The squeezing of the magnon mode is represented by $\mathcal{H}_{{MPA}}$, where $\lambda$ and $\theta$ are the squeezing parameter and the phase, respectively. These parameters indicate the degree of squeezing and the phase angle associated with the magnon mode in the quantum system under consideration \cite{10}. In this study, we examine high-quality YIG (yttrium iron garnet) spheres with a diameter of 250 $\mu$m, consisting of ferric ions (Fe$^{3+}$) at a density of $\rho = 4.22 \times 10^{27} \, \text{m}^{-3}$. This configuration results in a total spin $S = 5/3\rho V = 7.07 \times 10^{14}$, where $S$ is the collective spin operator that follows the commutation relation $[S_\alpha, S_\beta] = i \epsilon_{\alpha\beta\gamma} S_\gamma$.
Therefore, the Hamiltonian's total is denoted  
\begin{equation}\label{h}
\begin{aligned}
\mathcal{H} / \hbar&=\sum_{j=a, m, c}\omega_j j^{\dagger} j+\frac{{\omega_b}_1}{2}\left(q_1^2+p_1^2\right)+\frac{{\omega_b}_2}{2}\left(q_2^2+p_2^2\right)\\
&+ g_a\left(a^{\dagger} m+a m^{\dagger}\right)+g_m m^{\dagger} m q_1-g_1 c^{\dagger} c q_1\\
&-g_2 c^{\dagger} c q_2+i\lambda\left(e^{i\theta} {m^\dagger}^2 e^{-2i\omega_Lt}-e^{-i\theta} m^2e^{2i\omega_Lt}\right)\\
&+i \Omega\left(m^{\dagger} e^{-i \omega_L t}-\text { H.c }\right)+i \epsilon_L\left(c^{\dagger} e^{-i \omega_L t}-\text { H.c }\right)\\
&+i \epsilon_p\left(c^{\dagger} e^{-i \omega_p t}-\text { H.c }\right).
\end{aligned}
\end{equation}
The Hamiltonian \eqref{h} can be used to express the quantum Heisenberg-Langevin equations in a frame that rotates at the frequency $\omega_L$ as
\begin{equation}
	\begin{aligned}
	    &\dot{c}=-(\kappa_c+i\Delta_c)c+i g_1 c q_1+i g_2 c q_2+\epsilon_L\\
		&+\epsilon_p e^{-i\delta t}+\sqrt{2 \kappa_c} c_{i n}\\
		& \dot{a}=-(\kappa_a+i\Delta_a) a-i g_a m+\sqrt{2 \kappa_a} a_{i n} \\
		& \dot{m}=-(\kappa_m+i \Delta_m)m-i g_a a-i g_m m q_1+2\lambda e^{i\theta}m^\dagger\\
		&+\Omega+\sqrt{2 \kappa_m} m_{i n} \\
		& \dot{q_1}={\omega_b}_1 p_1,\:  \dot{p_1}=-{\omega_b}_1 q_1-{\gamma_b}_1 p_1+g_1 c^{\dagger} c-g_m m^{\dagger} m+\xi_1 \\
		& \dot{q_2}={\omega_b}_2 p_2,\: \dot{p_2}=-{\omega_b}_2 q_2-{\gamma_b}_2 p_2+g_2 c^{\dagger} c+\xi_2,  
	\end{aligned}
\end{equation}
where $\Delta_c=\omega_c-\omega_L$, $\Delta_a=\omega_a-\omega_L$, $\Delta_m=\omega_m-\omega_L$ and $\delta=\omega_p-\omega_L$ are detuning parameters. $\kappa_j$ $(j=c,a,m)$, ${\gamma_b}_1$ and ${\gamma_b}_1$ being the dissipation rates of the optical cavity, microwave cavity, magnon, mechanical of the mirror $q_1$ and mechanical of the mirror $q_2$ modes, respectively. The terms $j_{int}$ are the vacuum input noise operators, which have zero mean values. $\xi_1$ and $\xi_2$ are the Hermitian Brownian noise operator for mechanical oscillators.\\
By concentrating on the steady-state values and examining only the first-order terms in the fluctuating operator, the quantum Langevin equations \eqref{h} can be linearized. This can be expressed as
\[
\langle \mathcal{Z} \rangle = \mathcal{Z}_s + \mathcal{Z}_- e^{-i\delta t} + \mathcal{Z}_+ e^{i\delta t}.
\]
Here \(\mathcal{Z} = c, a, m,q_1, p_1, q_2, p_2\). In the first instance, we analyze steady-state solutions, also known as zero-order solutions, (see Appendix \ref{B}).\\
The solution for the cavity mode can be expressed as (see Appendix \ref{A})
\begin{equation}
	c_-=\epsilon_p\left[\alpha_1+\frac{G_2^2}{\mathcal{E}}+\frac{G_1^2\mathcal{M}}{\mathcal{N}}\left(1+\frac{G_2^2}{\mathcal{E}\alpha_2}\right)\right]^{-1}.
\end{equation}
By applying the standard input-output relation, given by $\epsilon_{\text{out}} = \epsilon_{\text{int}} - 2\kappa_c \langle c \rangle$ \cite{13}, we can represent the amplitude of the output field as follows
\begin{equation} \label{Y}
	\epsilon_{\text{out}} = \frac{2\kappa_c c_-}{\epsilon_p}.
\end{equation}
The real part $Re[\epsilon_{\text{out}}]$ of Eq. \eqref{Y} represents the absorption spectrum, while the imaginary part $Im[\epsilon_{\text{out}}]$ corresponds to the dispersion spectrum of the output field at the probe frequency.

\section{MAGNOMECHANICALLY INDUCED TRANSPARENCY} \label{I}
In this section, we numerically investigate the impact of the photon-phonon coupling strengths $G_{c1}$ and $G_{c2}$, as well as the microwave-magnon coupling strength $g_a$, on magnomechanically induced transparency within a cavity-OMM system. We investigate the impact of the squeezing parameter $\lambda$ on the transparency window. We utilize the effective parameters derived from a recent experiment in the cavity magnomechanical system, such as \cite{32,44,52}: $\omega_{c} / 2 \pi=10$ $\mathrm{GHz}$, $\omega_b / 2 \pi=\omega_{b1} / 2 \pi=\omega_{b2} / 2 \pi=10$ $\mathrm{MHz}, $ ${\gamma_b}_1 / 2 \pi={\gamma_b}_2 / 2 \pi=$ $100 \mathrm{~Hz}$, $\omega_{m} / 2 \pi=10$ $\mathrm{GHz},$ $\kappa_{c} / 2 \pi=2$ $\mathrm{MHz},$ $\kappa_a / 2 \pi=1.5$ $\mathrm{MHz},$ $\kappa_{m} / 2 \pi=1.5$ $\mathrm{MHz},$ $G_{c1} / 2 \pi=3.2$ $\mathrm{MHz},$ $G_{c2} / 2 \pi=4.8$ $\mathrm{MHz},$ $G_m / 2 \pi=4.8$ $\mathrm{MHz},$ $g_{a} / 2 \pi=$ 2 $\mathrm{MHz}, \tilde{\Delta}_c=\omega_b,$ and $\tilde{\Delta}_{m}=\Delta_a=-\omega_b$.\\ 
\begin{figure} [h!] 
	\begin{center}
		\includegraphics[scale=0.21]{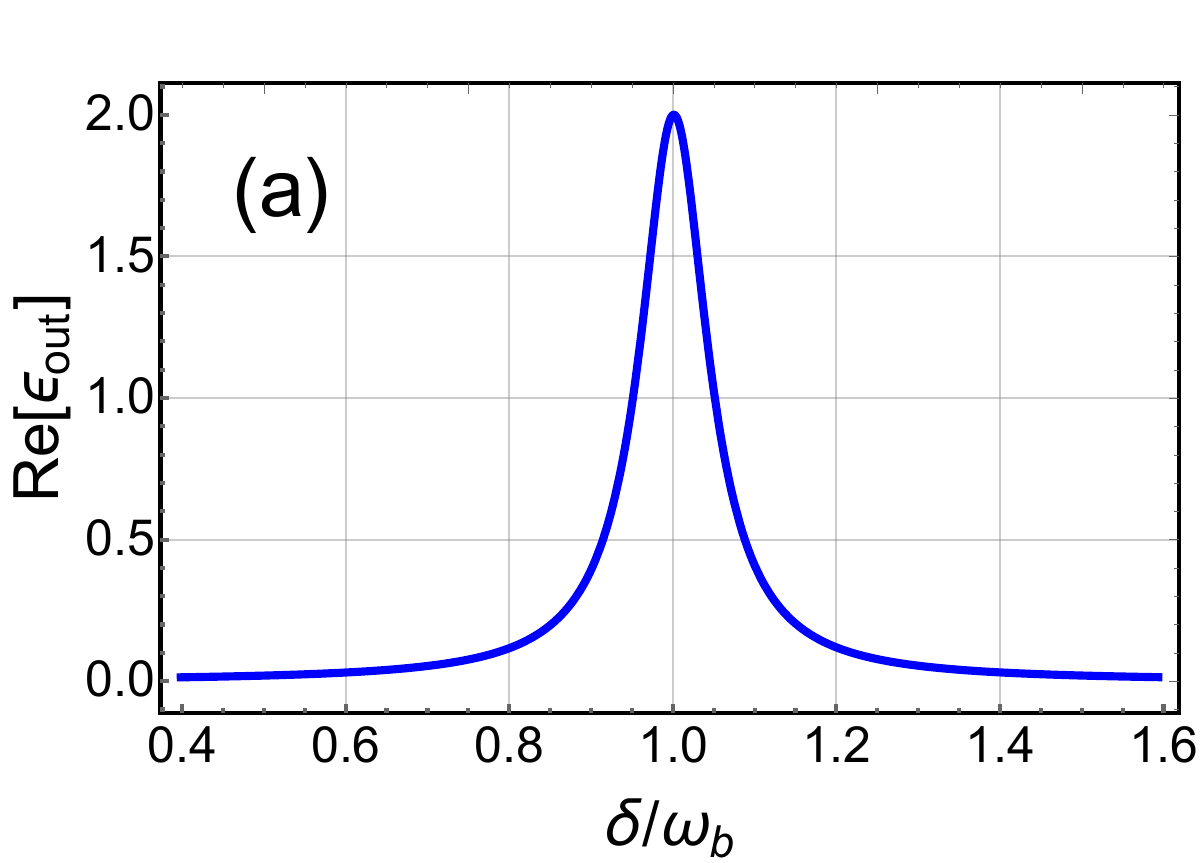}
		\includegraphics[scale=0.21]{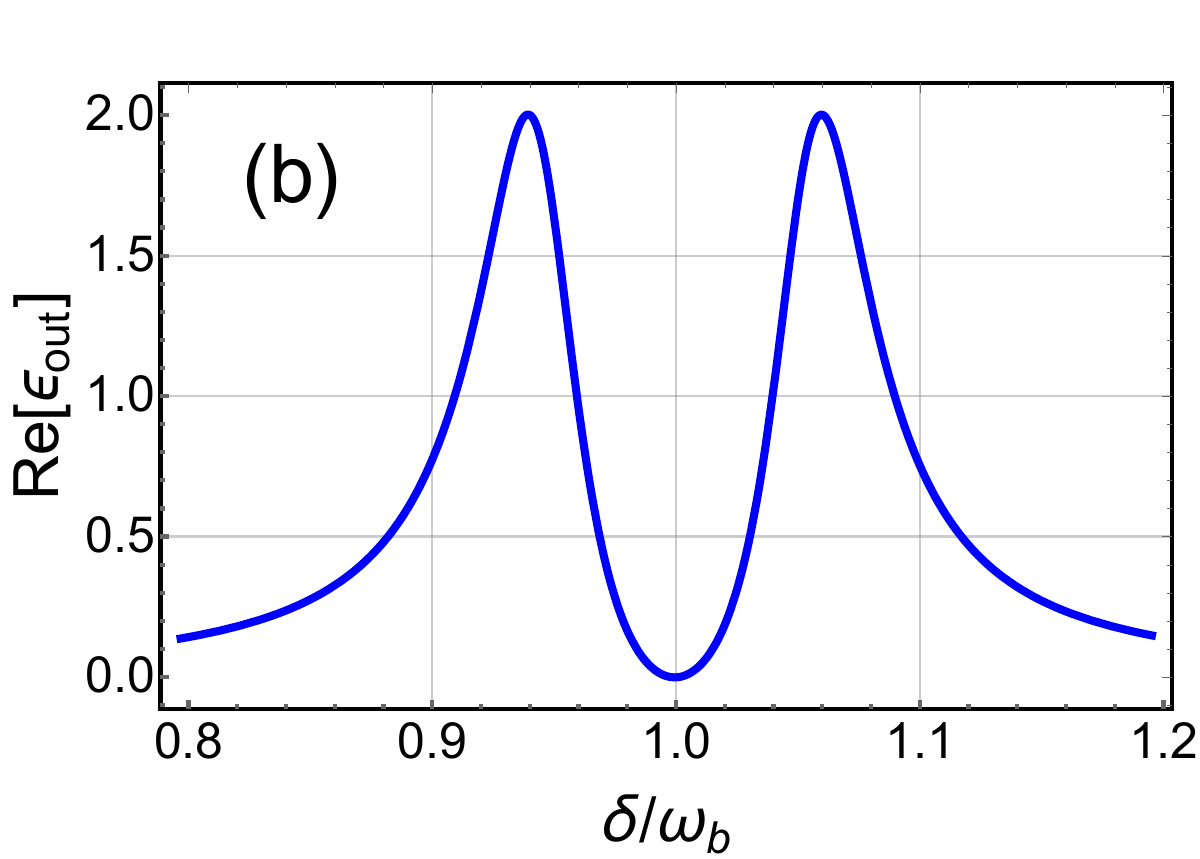}
		\includegraphics[scale=0.21]{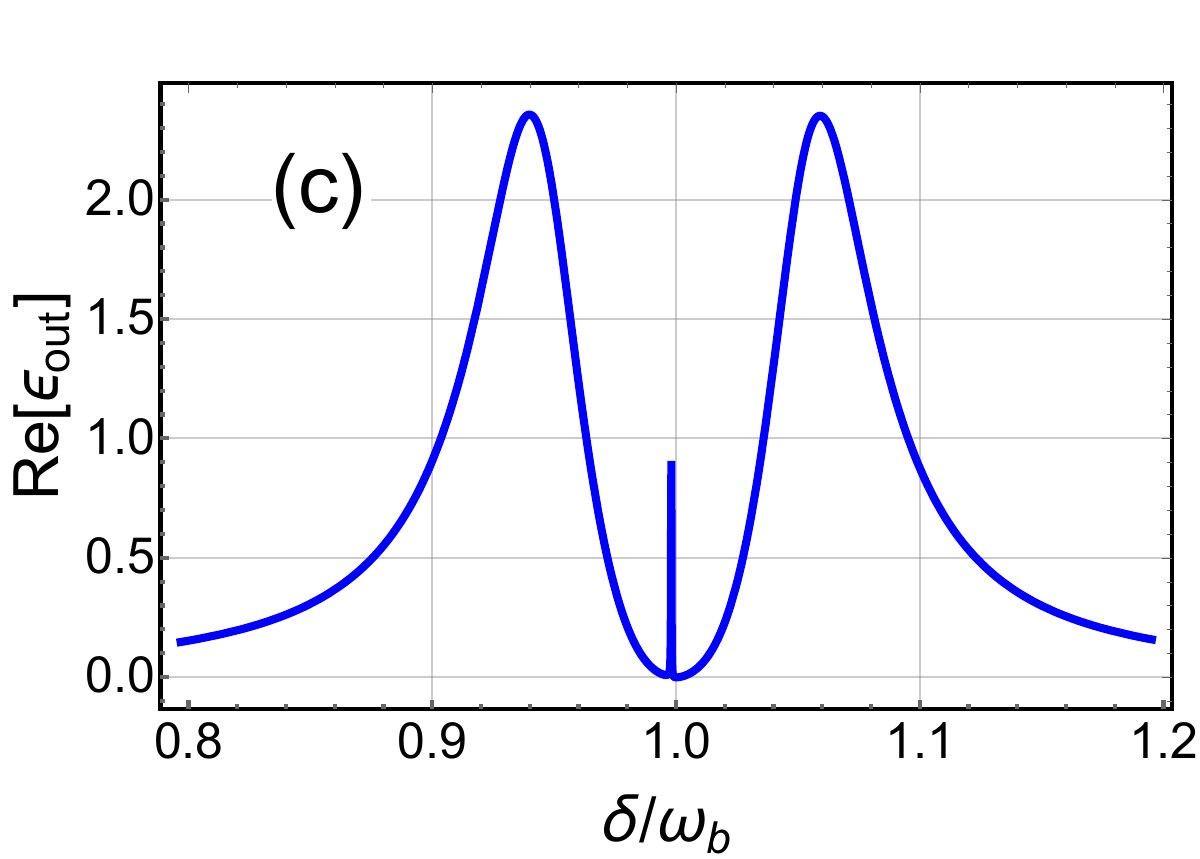}
		\includegraphics[scale=0.21]{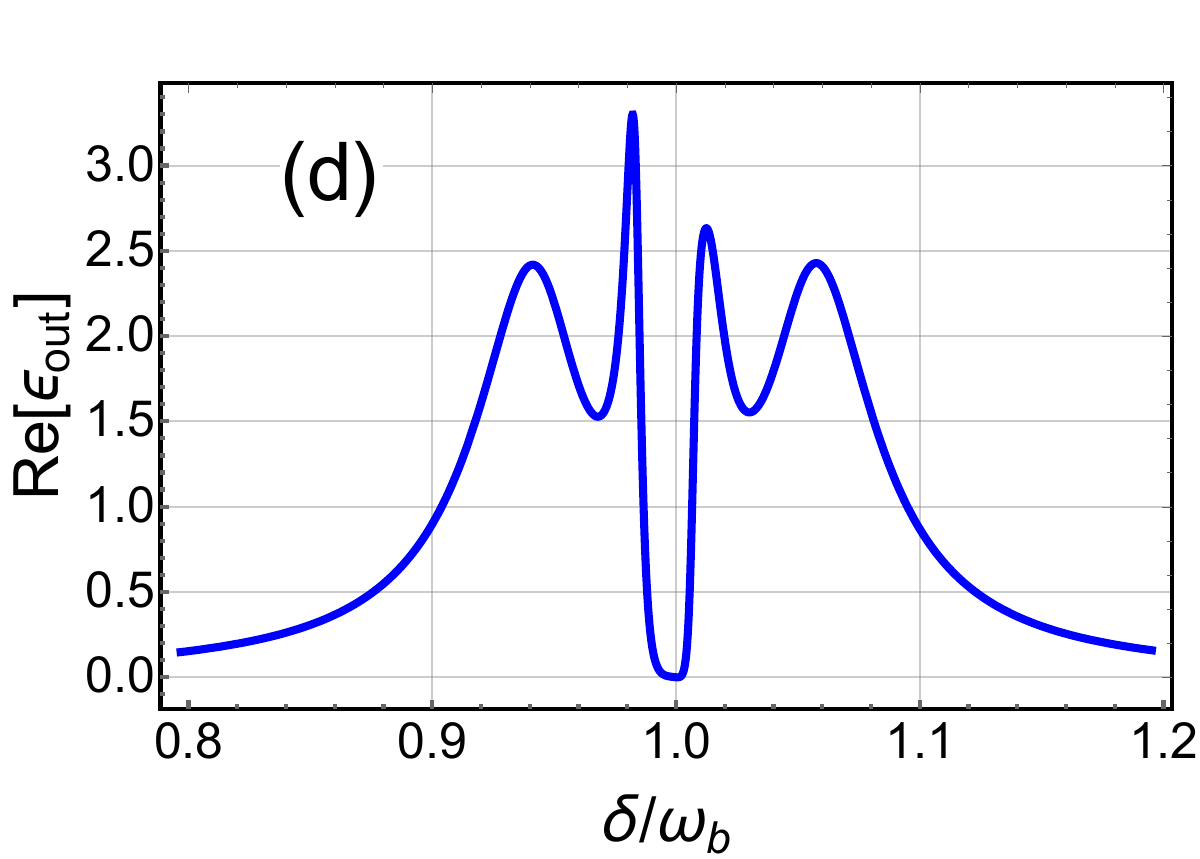}
		\includegraphics[scale=0.21]{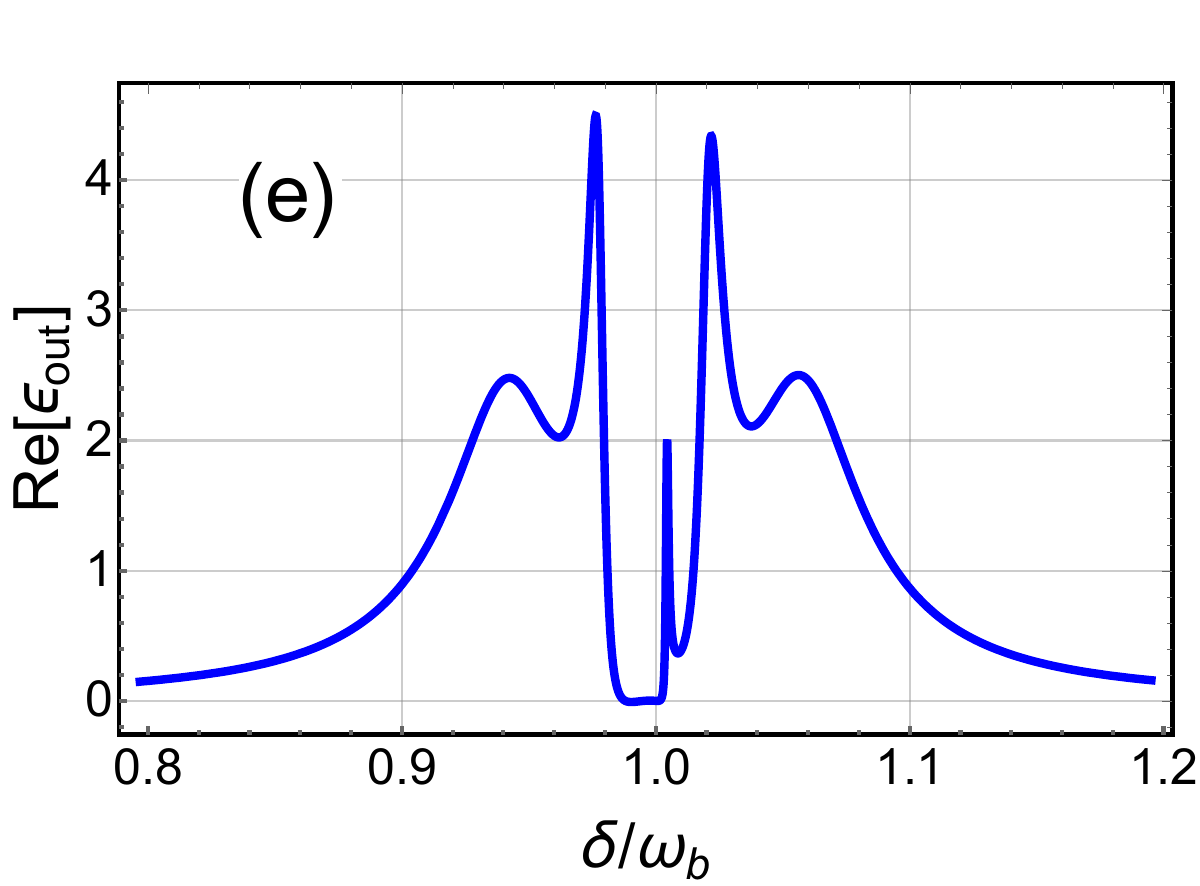}
		\caption{Plot of the real Re[$\epsilon_{out}$] part of the output field as a function of the normalized detuning $\delta/\omega_{b}$ for different values of the coupling strength. (a) $G_{c1}=G_{c2}=g_a=0$, (b) $G_{c1}=g_a=0$, $G_{c2}/2\pi=4.8$ MHz, (c) $G_{c1}/2\pi=3.2$ MHz, $G_{c2}/2\pi=4.8$ MHz and $g_a=0$ MHz, (d) $G_{c1}/2\pi=3.2$ MHz, $G_{c2}/2\pi=4.8$ MHz and $g_a/2\pi=2$ MHz, and (e) $G_{c1}/2\pi=3.2$ MHz, $G_{c2}/2\pi=4.8$ MHz and $g_a/2\pi=2.4$ MHz. In all panels, $\lambda=0.5\kappa_{c}$, $\theta=0$, and see the text for other parameters.} \label{b}
	\end{center}
\end{figure} 
In Figs. \ref{b}(a)-\ref{b}(b), we plot the absorption spectrum of the output field versus the normalized probe detuning $\delta/\omega_d$ for different coupling strengths. We observe in Fig. \ref{b}(a) that there are no signatures of transparency windows in the absorption spectrum when the photon mode is not coupled with the phonon couplings ($G_{c1}=0$ and $G_{c2}=0$) and the interaction between the magnon mode and the microwave cavity is absent ($g_a=0$). In Fig. \ref{b}(b), we activated the second photon-phonon coupling $G_{c2}$ while keeping the first photon-phonon coupling $G_{c1}$ and the microwave-magnon coupling $g_a$ absent. With these conditions, a single transparency window appears in the output probe field, characterized by two peaks and a single dip. Fig. \ref{b}(c) shows that there are two transparency windows when the first photon-phonon coupling is activated while the microwave-magnon coupling remains absent, i.e., $G_{c2}\neq0$, $G_{c2}\neq0$, and $g_a = 0$. It is observed that the number of transparency windows increases from one in Fig. \ref{b}(b) to two in Fig. \ref{b}(c), as a result of the introduction of mechanical oscillators $G_{c1}$. In Fig. \ref{b}(d), it can be noticed that the absorption spectrum of the probe field exhibits three transparency windows when $G_{c1}$, $G_{c2}$, and $g_{a}$ are present. Compared to Fig. \ref{b}(d), an additional transparency window appears due to the microwave-magnon coupling strength $g_a/2\pi=2$ MHz. Therefore, the interaction between microwave mode and magnon mode leads to an increase in the number of transparency windows. In Fig. \ref{b}(e), when we fixed the microwave-magnon coupling strength in $g_a/2\pi=2.4$ MHz, we observe four transparency window, which composed of five peaks and four dips. Obviously, compared with Figure \ref{b}(d), the number of the transparency windows is increased by increasing the microwave-magnon interaction.\\
\begin{figure} [h!] 
	\begin{center}
		\includegraphics[scale=0.21]{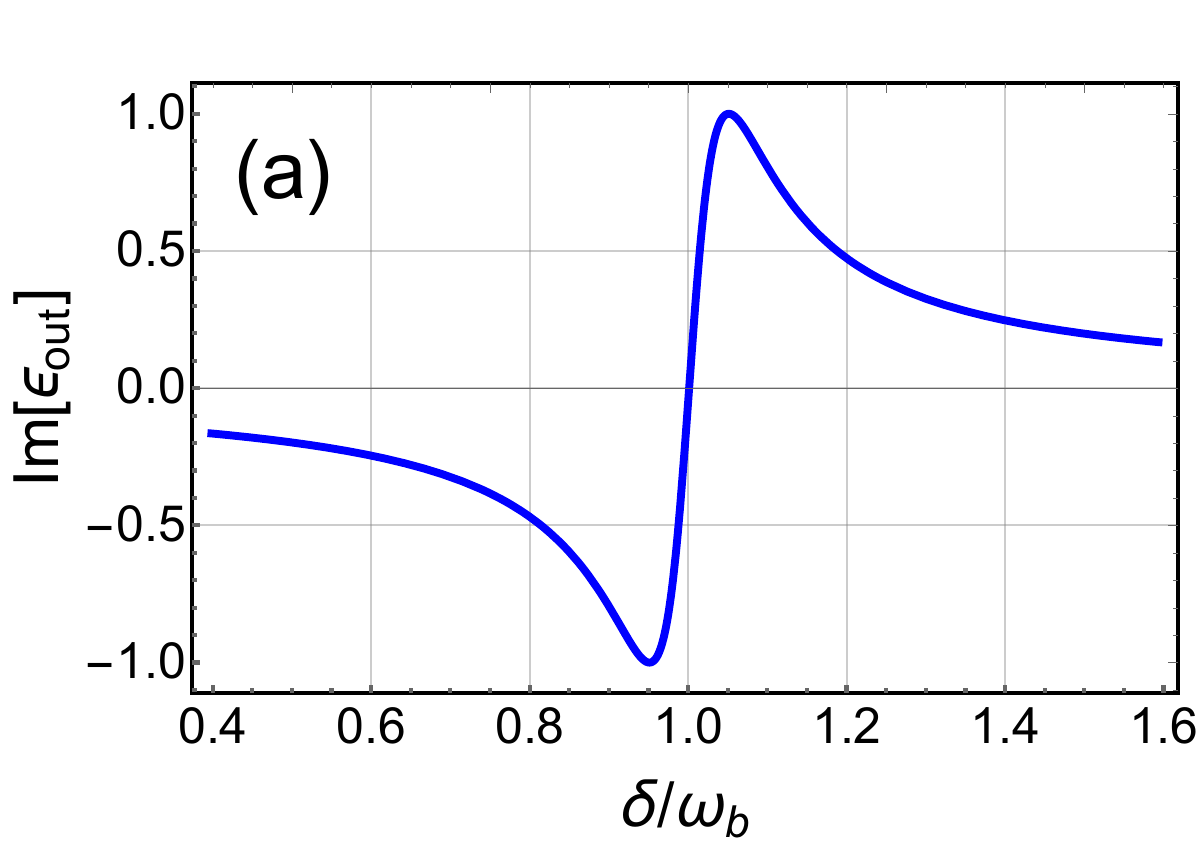}
		\includegraphics[scale=0.21]{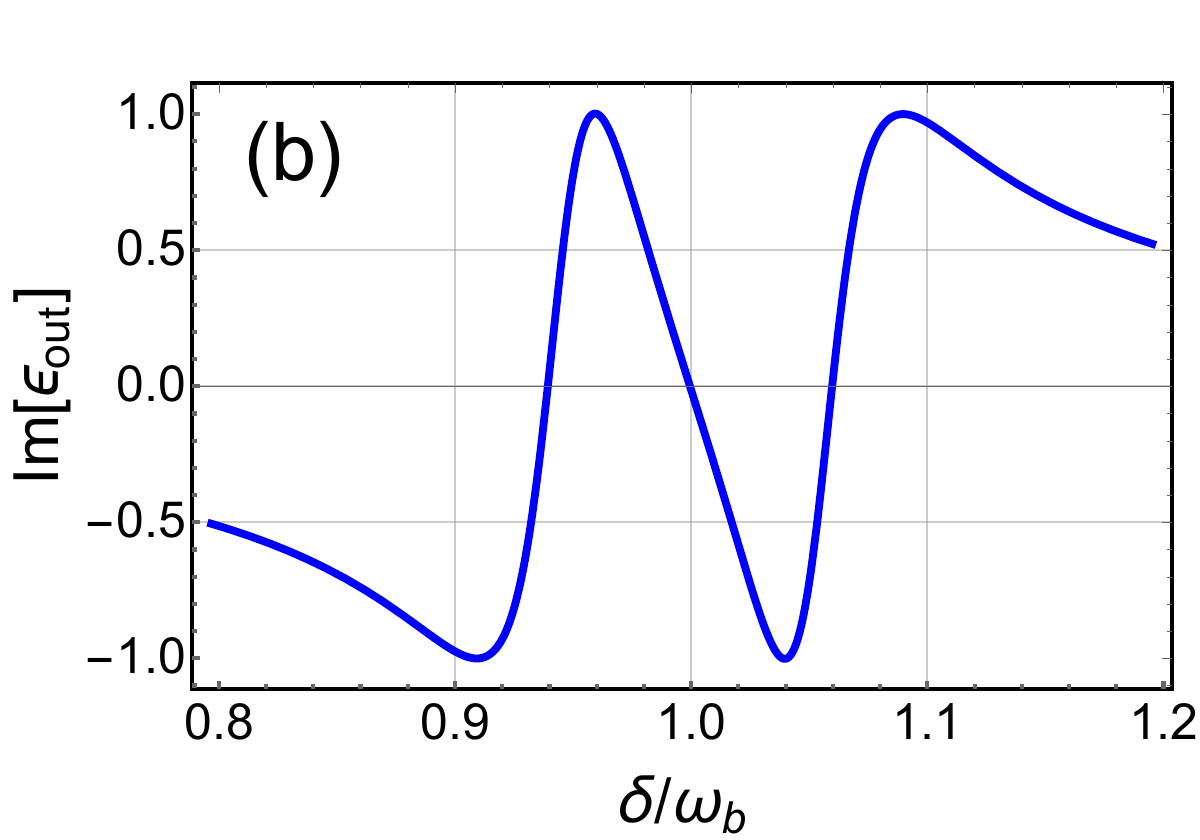}
		\includegraphics[scale=0.21]{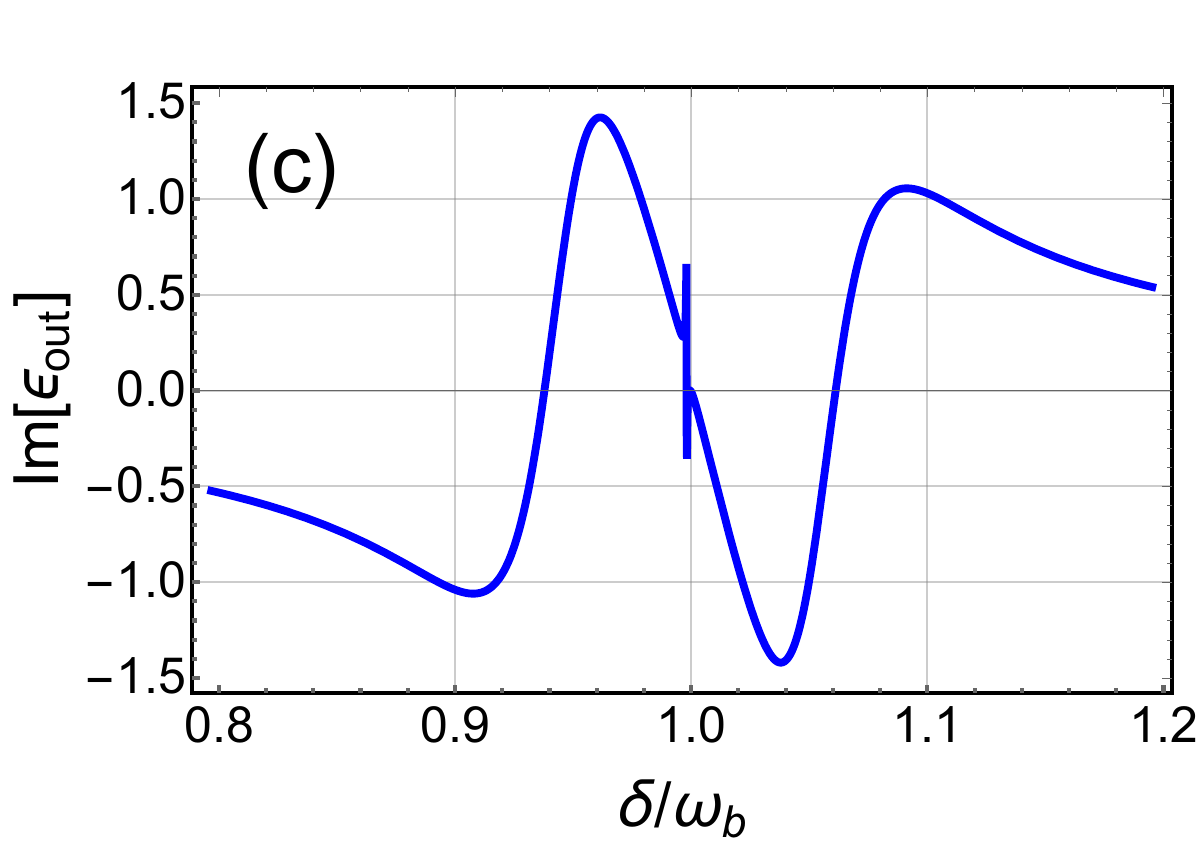}
		\includegraphics[scale=0.21]{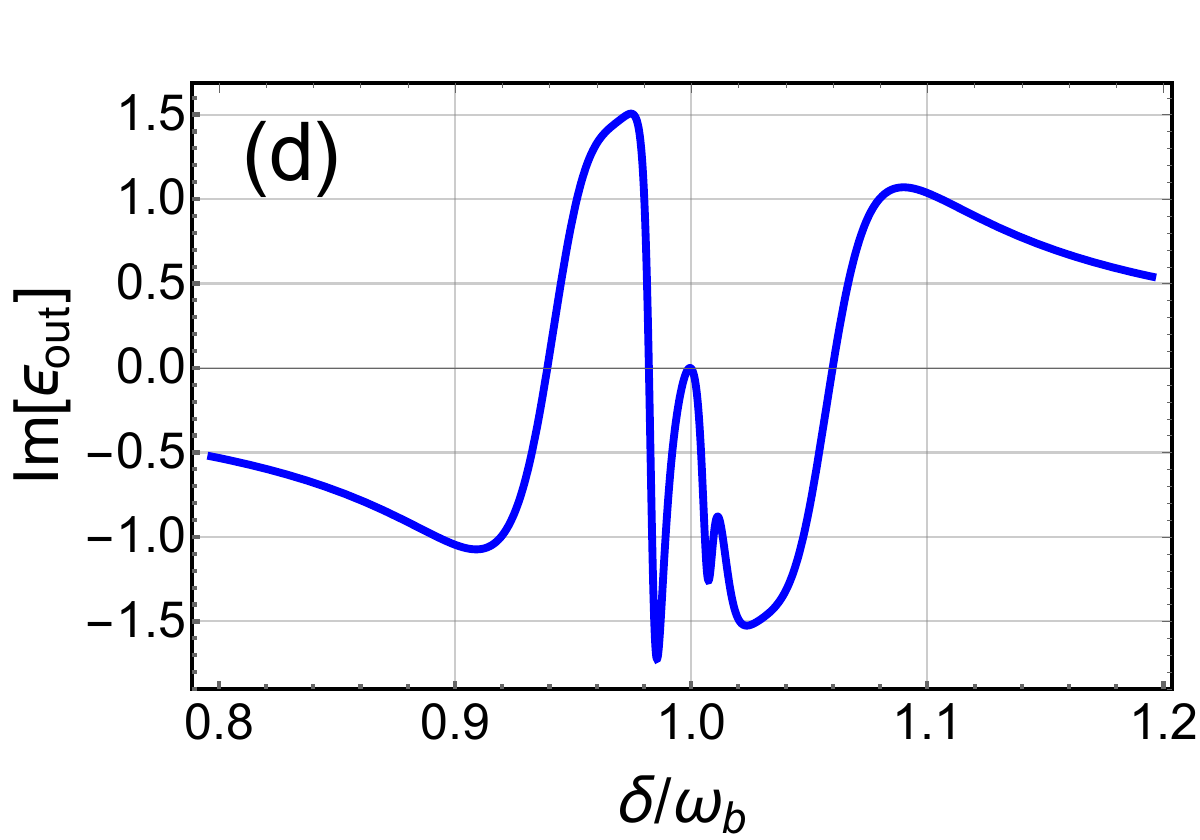}
		\includegraphics[scale=0.21]{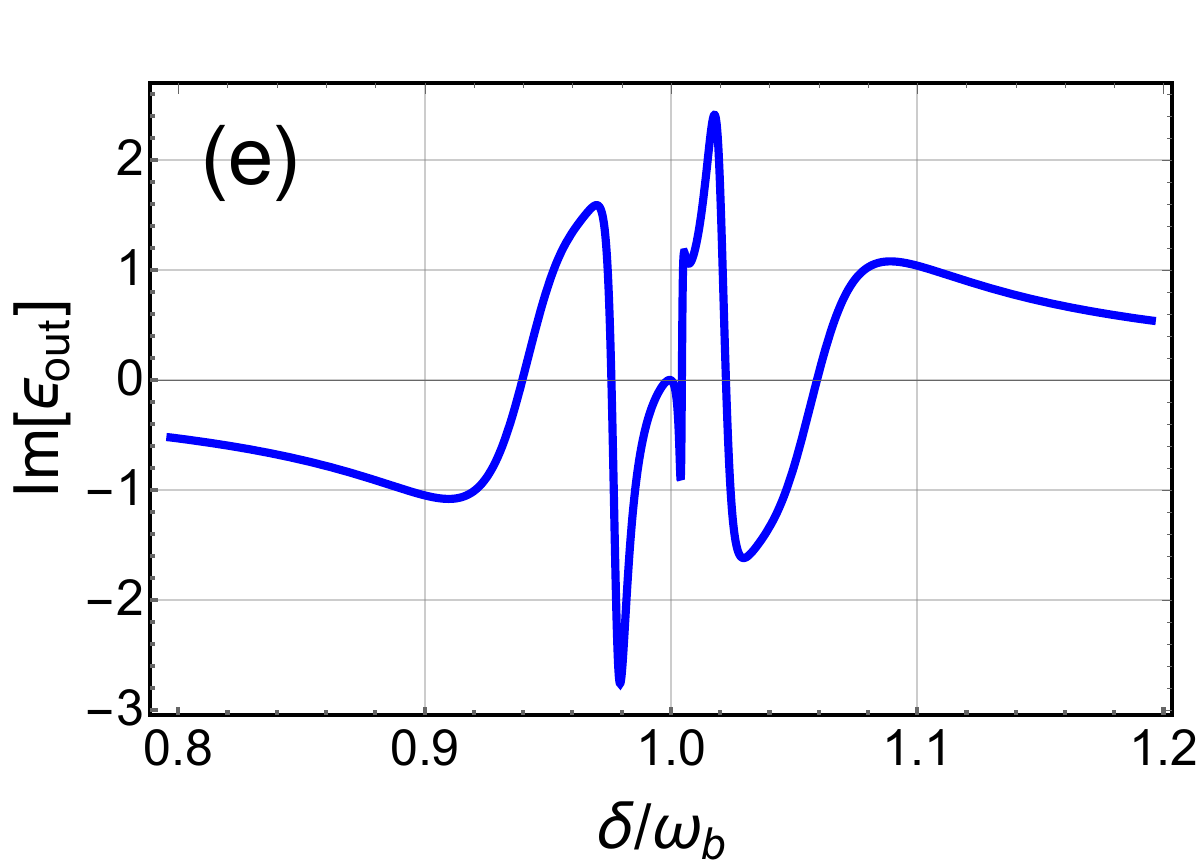}
		\caption{Plot of the imaginary Im[$\epsilon_{out}$] part of the output field as a function of the normalized detuning $\delta/\omega_{b}$ for different values of the coupling strength. (a) $G_{c1}=G_{c2}=g_a=0$, (b) $G_{c1}=g_a=0$, $G_{c2}/2\pi=4.8$ MHz, (c) $G_{c1}/2\pi=3.2$ MHz, $G_{c2}/2\pi=4.8$ MHz and $g_a=0$ MHz, (d) $G_{c1}/2\pi=3.2$ MHz, $G_{c2}/2\pi=4.8$ MHz and $g_a/2\pi=2$ MHz, and (e) $G_{c1}/2\pi=3.2$ MHz, $G_{c2}/2\pi=4.8$ MHz and $g_a/2\pi=2.4$ MHz. In all panels, $\lambda=0.5\kappa_{c}$, $\theta=0$, and see the text for other parameters.} \label{1b}
	\end{center}
\end{figure}
In Figs. \ref{1b}(a)-\ref{1b}(e), we plot the dispersion spectrum of the transmitted field as a function of the scaled detuning $\delta/\omega_{b}$ for several values of coupling strengths. Fig. \ref{1b}(a) clearly shows that no signatures of MMIT appear in the dispersion spectrum of the probe field when the optical mode is not coupled with the mechanical mode $(G_{c1}=G_{c2}=0)$ and the microwave mode is note coupled with the magnon mode $(g_a=0)$. In Fig. \ref{1b}(b), with the activation of photon-phonon coupling $G_{c2}$, while keeping $G_{c1}=0$ and $g_a=0$, single transparency windows is observed in the dispersion spectrum. Next, we introduce photon–phonon coupling $G_{c2}$ in Fig. \ref{1b}(c). The single transparency windows splits into double windows. Figs. \ref{1b}(d) and \ref{1b}(e) display the dispersion spectrum of the output field when all three couplings are simultaneously present. We remark three and four transparency windows in the dispersion spectrum for $g_a/2\pi=2$ MHz and $g_a/2\pi=2.4$ MHz, respectively. We can therefore conclude that the interaction between the microwave mode and the magnon mode constitutes two transparency windows.\\
\begin{figure} [h!] 
	\begin{center}
		\includegraphics[scale=0.75]{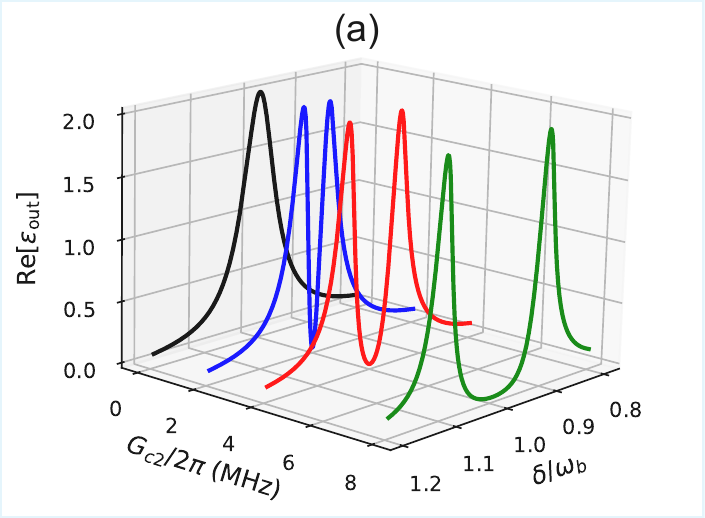}
		\includegraphics[scale=0.28]{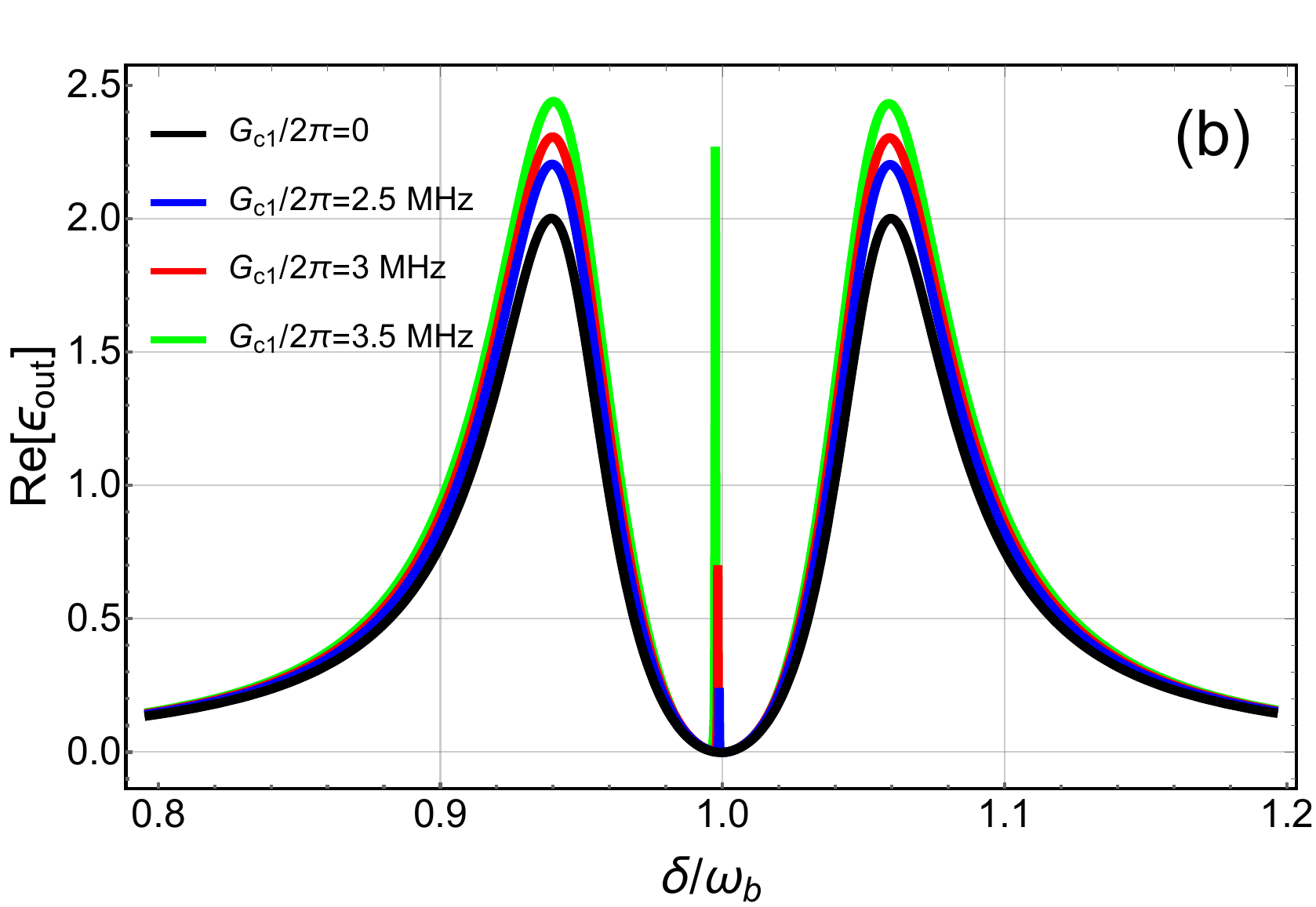}
		\caption{Plot of the real Re[$\epsilon_{out}$] part of the output field as a function of the normalized detuning $\delta/\omega_{b}$ for different values of the optomechanical couplings strength $G_{c2}$ and $G_{c1}$; (a) $G_{c2}=0$, $G_{c2}/2\pi= 2$ MHz, $G_{c2}/2\pi= 4$ MHz and $G_{c2}/2\pi= 8$ MHz with $G_{c1}=0$, $G_{m}/2\pi=4.8$ MHz, $g_{a}= 0$, $\lambda=0.5\kappa_{c}$ and $\theta=0$; and (b) $G_{c1}=0$, $G_{c1}/2\pi= 2$ MHz, $G_{c1}/2\pi= 3.1$ MHz, $G_{c1}/2\pi= 3.2$ MHz and $G_{c2}/2\pi= 3.3$ MHz with $G_{c2}/2\pi= 4.8$ MHz, $G_{m}/2\pi=4.8$ MHz, $g_{a}= 0$, $\lambda=0.5\kappa_{c}$ and $\theta=0$. } \label{bb}
	\end{center}
\end{figure} 
In Fig. \ref{bb}(a), we show the absorption spectrum Re[$\epsilon_{out}$] of the output field for different values of the optomechanical coupling $G_{c2}$ versus normalized detuning of the probe field via keeping the $G_{c1}=0$, ${G_m}=2\pi\times4.8$ MHz, $\lambda=0.5\kappa_{c}$ and ${g_a}=0$. We consider the four cases with $G_{c2}= 0$, $G_{c2}/2\pi= 2$ MHz, $G_{c2}/2\pi= 4$ MHz and $G_{c2}/2\pi= 8$ MHz. The black line, corresponding to the case where $G_{c2}= 0$, is already discussed in Fig. \ref{b}(a).
The blue, red and green lines show that increasing the photon-phonon coupling $G_{c2}$ widens the transparency window. This suggests that the optomechanically induced transparency effect increases with increasing $G_{c2}$. Fig. \ref{bb}(b) illustrates the absorption transparency behavior of the output probe field in the presence of $G_{c2}$ for different values of the photon-phonon coupling $G_{c1}$;  $G_{c1}=0$, $G_{c1}/2\pi= 2$ MHz, $G_{c1}/2\pi= 3.1$ MHz, $G_{c1}/2\pi= 3.2$ MHz and $G_{c2}/2\pi= 3.3$ MHz. The case with $G_{c2} = 0$, shown by the black line, was previously examined in Fig. \ref{b}(b). When $G_{c1}/2\pi=3.1$ MHz, we can see that the transparency window increases from one in black line to two in the blue line. By comparing the blue, red and green, lines in Fig. \ref{bb}(b), it is evident that the width of the transparency window increases with higher values of the photon-phonon coupling $G_{c1}$.\\
\begin{figure} [h!] 
	\begin{center}
		\includegraphics[scale=0.4]{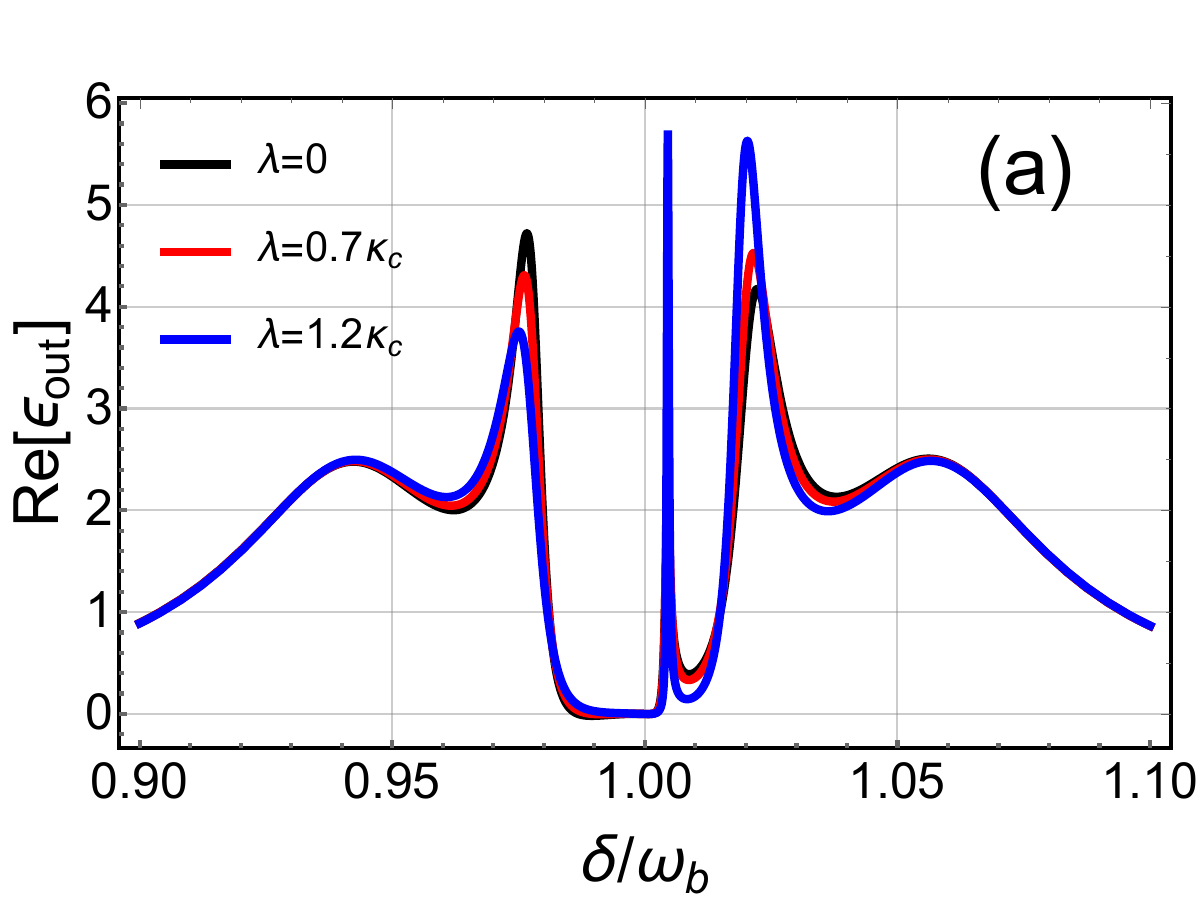}
		\includegraphics[scale=0.4]{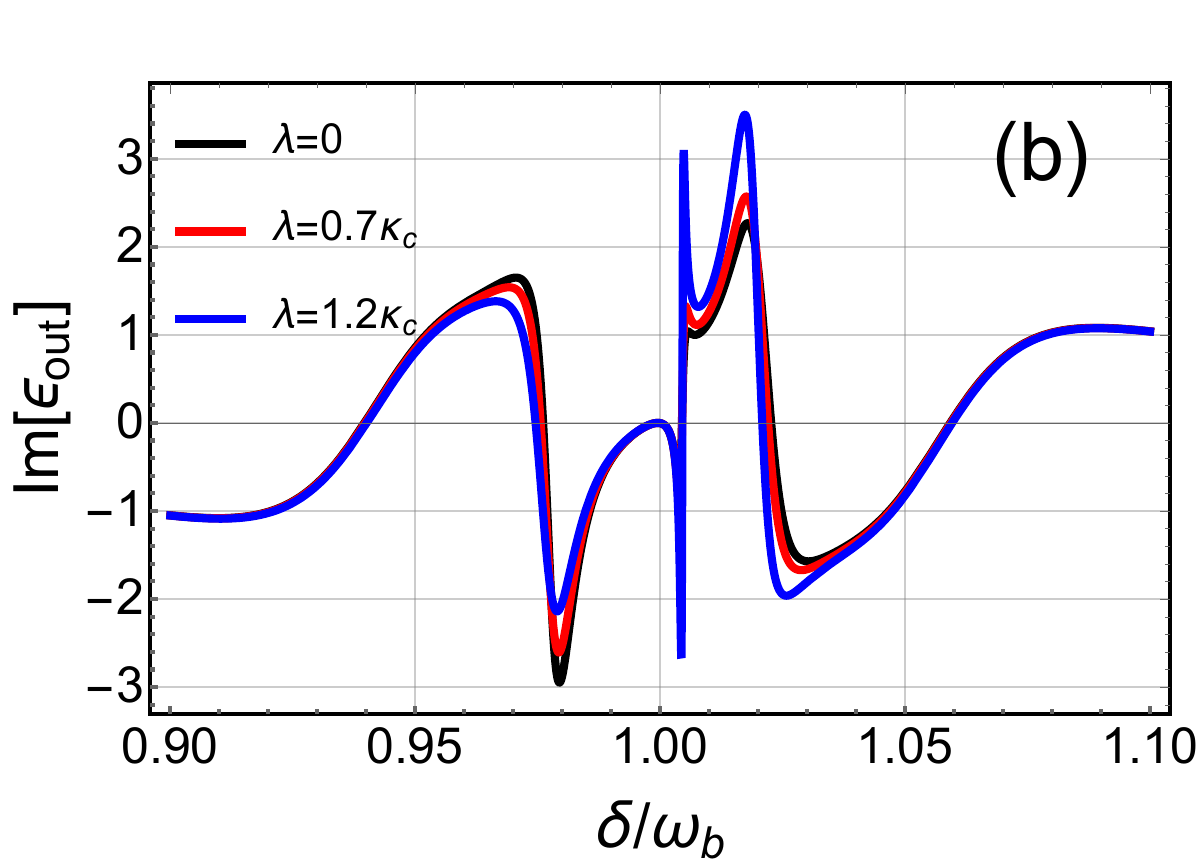}
		\caption{Plot of the real Re[$\epsilon_{out}$] and imaginary Im[$\epsilon_{out}$] parts of the output field as a function of the normalized detuning $\delta/\omega_{b}$ for different values of the squeezing parameter $\lambda$ with $G_{c1}/2\pi=3.2$ MHz, $G_{c2}/2\pi =G_{m}/2\pi=4.8$ MHz, $g_a/2\pi=2.4$ MHz and $\theta=0$. See the text for other parameters.} \label{X}
	\end{center}
\end{figure}
In Fig. \ref{X}, we present the absorption and the dispersion spectra as functions of the normalized probe field detuning $\delta/\omega_{b}$ for different squeezing parameter values $\lambda$, with the phase $\theta$ set to zero. In Fig. \ref{X}(a), we consider $G_{c1}= G_{m} = 2\pi\times3.2$ MHz and $g_{a} = 2\pi\times2.3$ MHz for various $\lambda$. The results show that increasing the parameter $\lambda$ leads to a narrower and deeper right transparency window, and a broader and shallower left transparency window. This indicates that the width of the transparent window can be actively controlled by tuning the degree of squeezing. In Fig. \ref{X}(b), we depict the output field dispersion for different squeezing parameter $\lambda$ versus the normalized detuning $\delta/\omega_{b}$. It is evident that the transparency window widens on one side and narrows on the other as the degree of squeezing increases.
\section{FANO RESONANCES} \label{II}
In this section, we will investigate the emergence of Fano line shapes in the output spectrum and examine the underlying physical mechanisms. Unlike the symmetric resonance profiles observed in phenomena such as EIT, OMIT, and MMIT windows \cite{25,27,28,31}, Fano resonances are distinguished by their notably asymmetric shapes. These resonances have been observed in systems that typically exhibit EIT, achieved through precise tuning of system parameters \cite{f1,f2}. The unique asymmetric profile of Fano resonances in systems with optomechanical-like interactions results from non-resonant interactions. For example, in a standard optomechanical setting, Fano resonance shapes in the spectral distribution can arise when the anti-Stokes process is detuned from the cavity's resonant frequency \cite{28,f1}.\\
\begin{figure} [h!] 
	\begin{center}
		\includegraphics[scale=0.4]{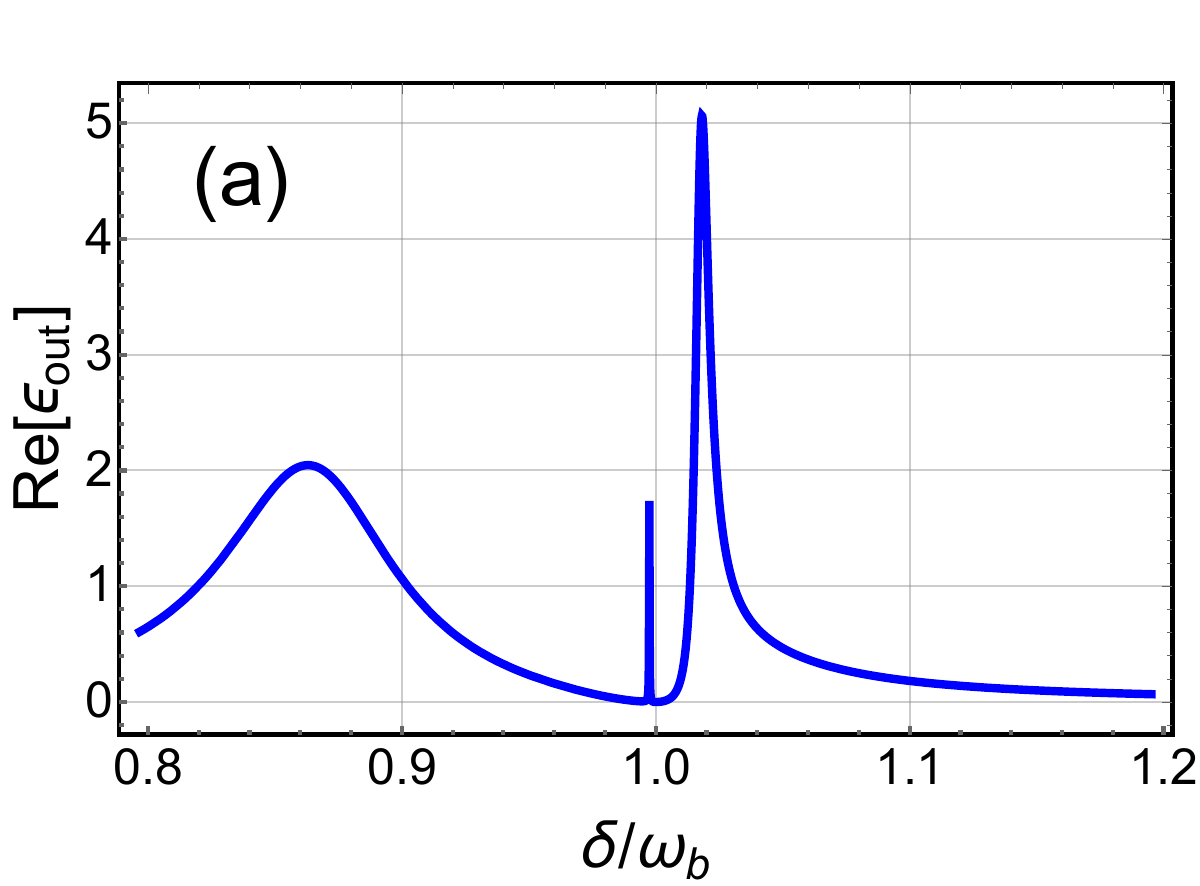}
		\includegraphics[scale=0.4]{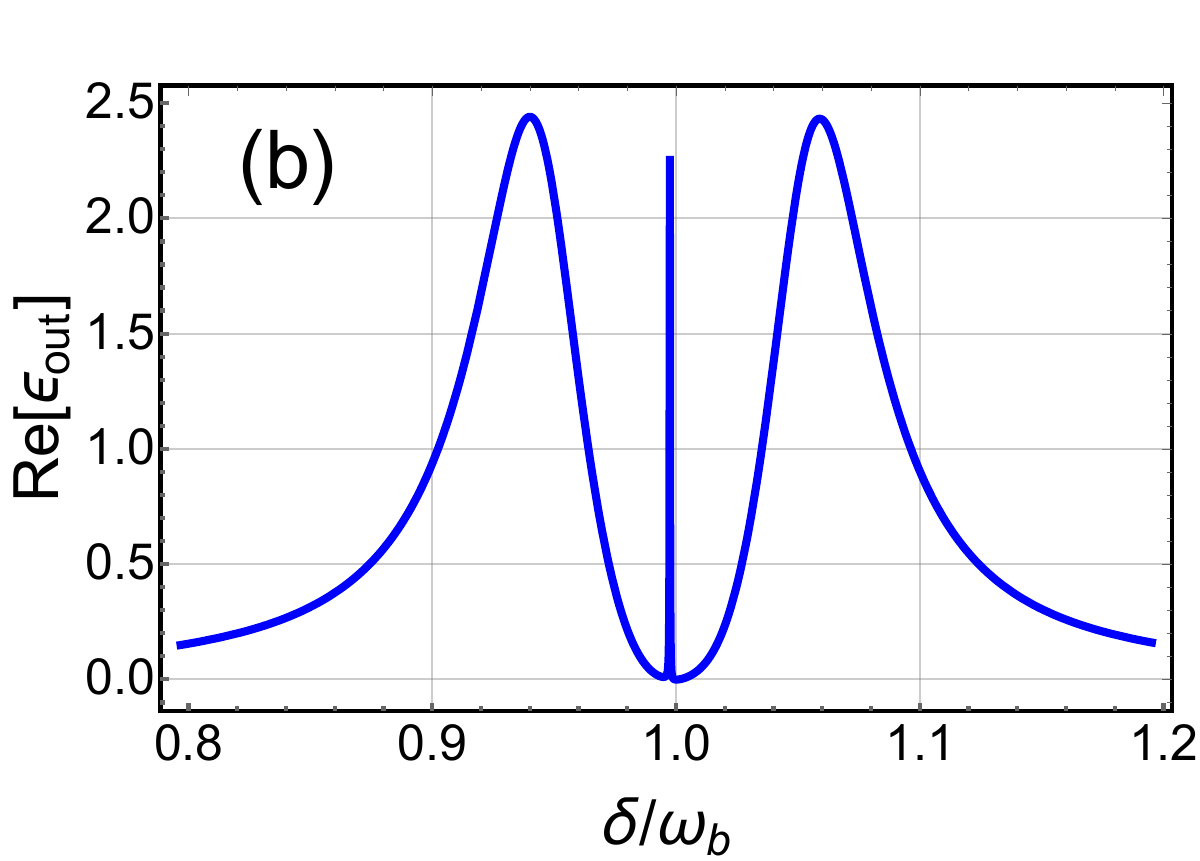}
		\caption{Fano lineshapes in the asymmetric absorption profiles are displayed versus the normalized probe frequency $ \delta / \omega_b $ with $G_{c1}/2\pi=3.5$ MHz, $G_{c2}/2\pi=4.8$ MHz, $G_{m}/2\pi= 4.8$ MHz, $g_{a}= 0$, $\lambda=0.5\kappa_{c}$ and $\theta=0$. (a) $\tilde{\Delta}_c=0.9\omega_{b}$ and (b) $\tilde{\Delta}_c=\omega_{b}$. The remaining parameters are provided in Sect. \ref{I}.}\label{F}
	\end{center}
\end{figure}
In Fig. \ref{F}(a)-\ref{F}(b), we plot the real part of the output field $\text{Re}[\epsilon_{out}]$ versus the normalized detuning $\delta/\omega_{b}$. In Fig. \ref{F}(a), with the phonon–photon coupling set to $G_{c1}/2\pi = 3.2$ MHz and $G_{c2}/2\pi = 4.8$ MHz, and the microwave-magnon coupling set to $g_a = 0$, the output spectrum displays an asymmetric window profile caused by a non-resonant process $(\tilde{\Delta}_c=0.9\omega_{b})$. In Fig. \ref{F}(b), the Fano resonance vanishes when considering the resonant case $\tilde{\Delta}_c = \omega_b$. These resonances occur due to the destructive interference between the weak probe field and the anti-Stokes scattering of light, which is induced by the mechanical oscillations resulting from the strong control field.
\section{SLOW AND FAST LIGHT}\label{III}
\begin{figure} [h!] 
	\begin{center}
		\includegraphics[scale=0.21]{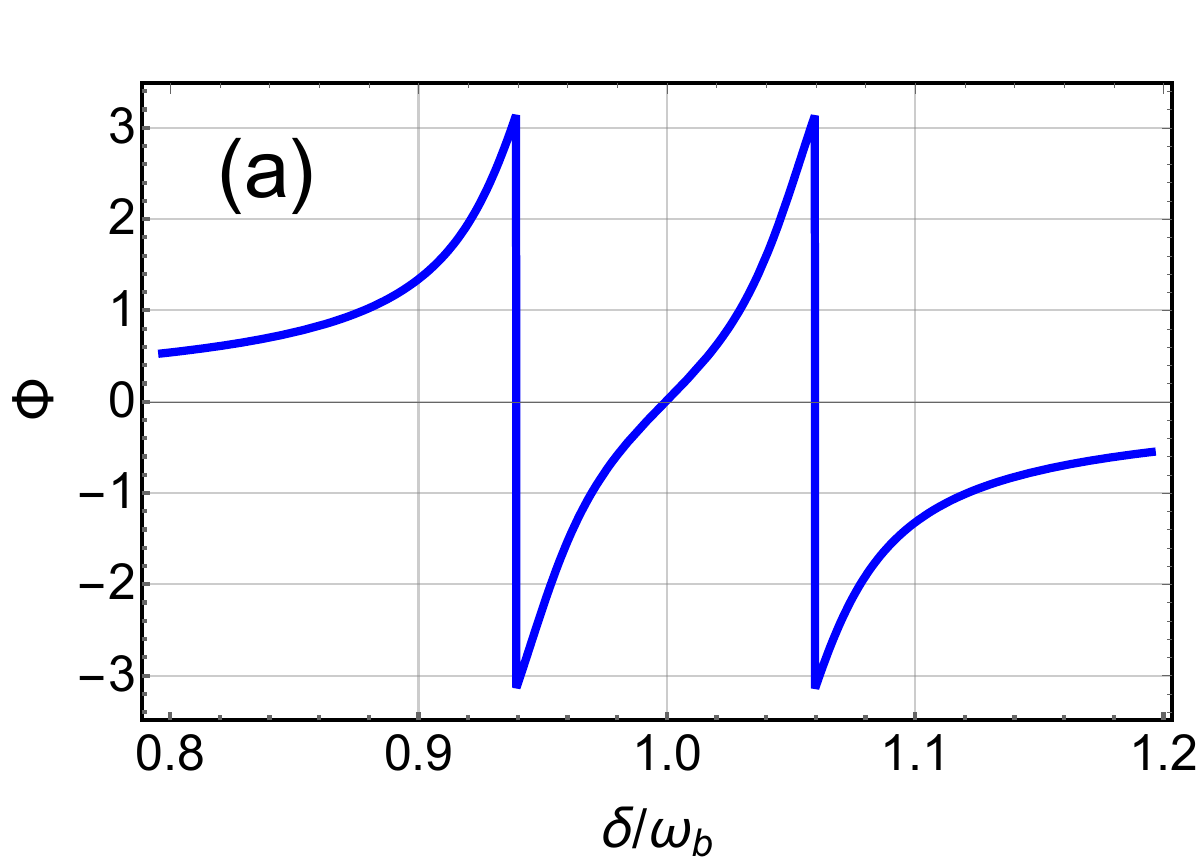}
		\includegraphics[scale=0.21]{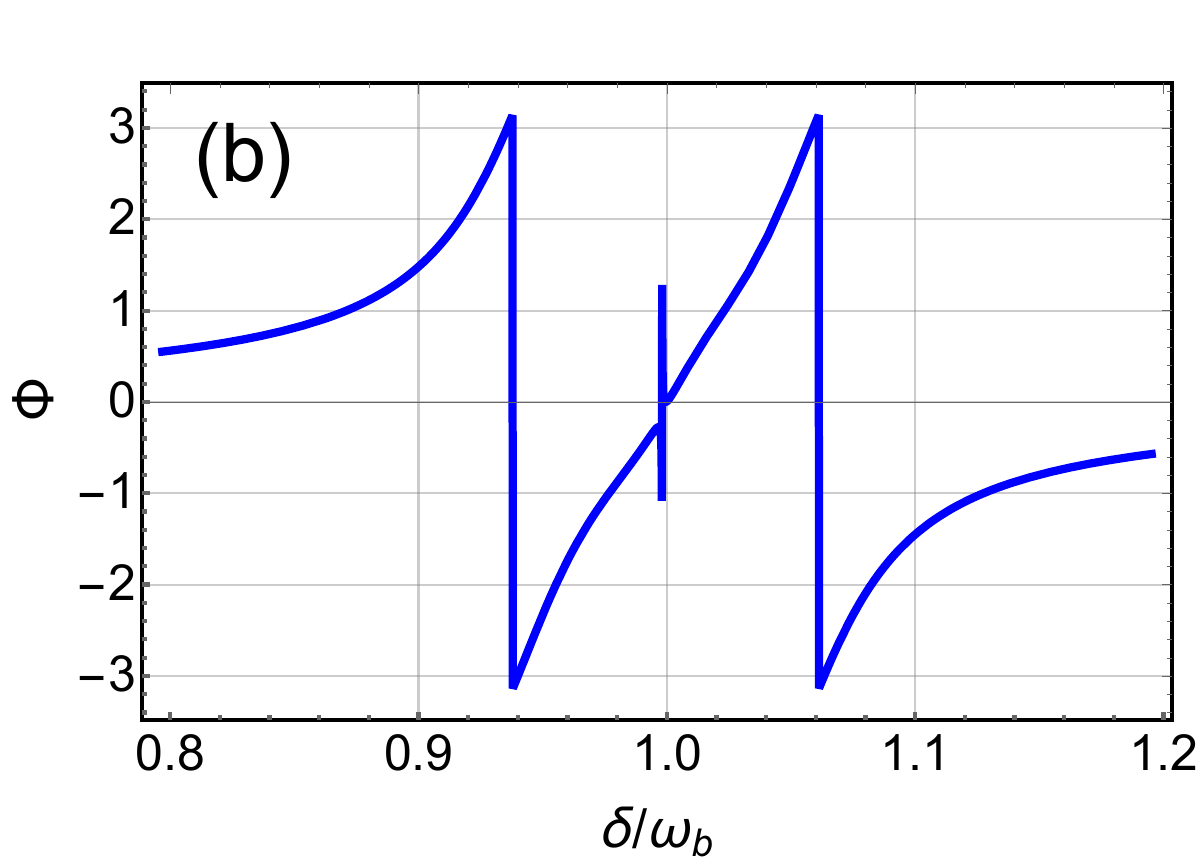}
		\includegraphics[scale=0.21]{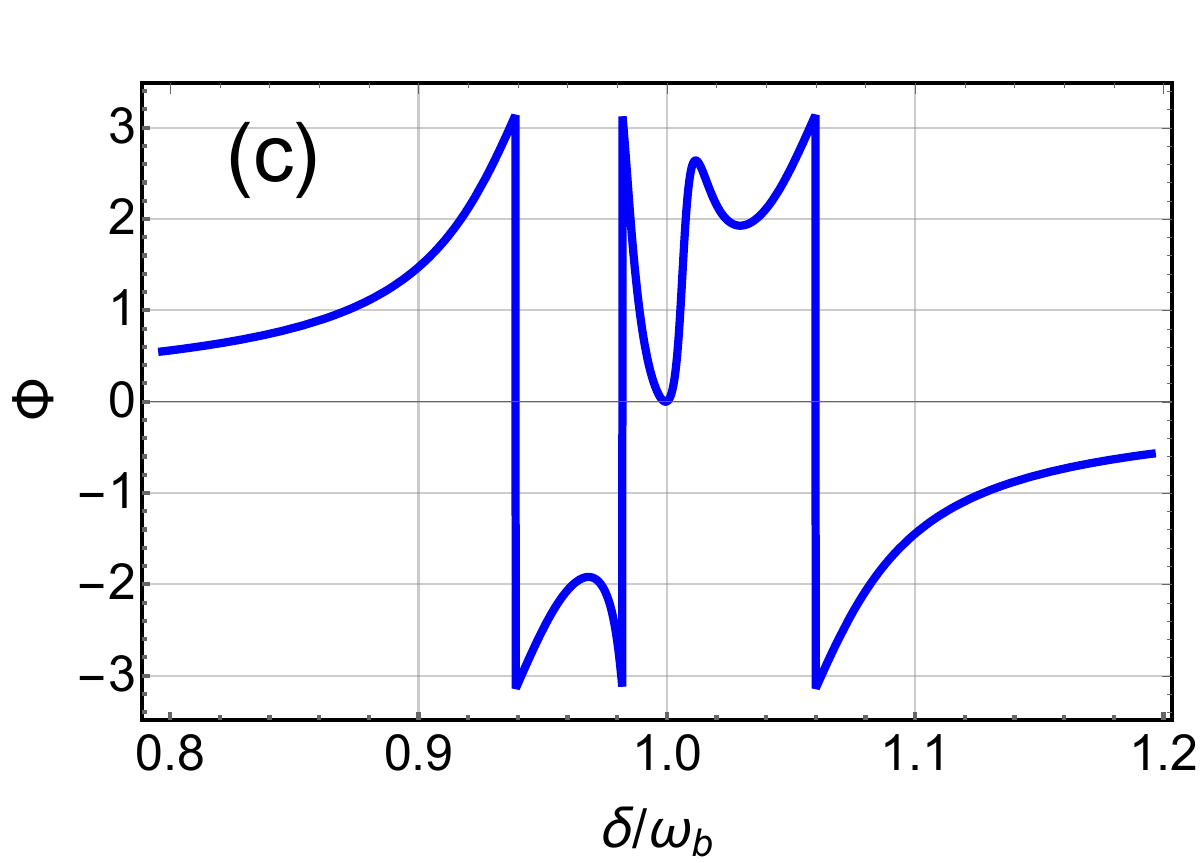}
		\includegraphics[scale=0.21]{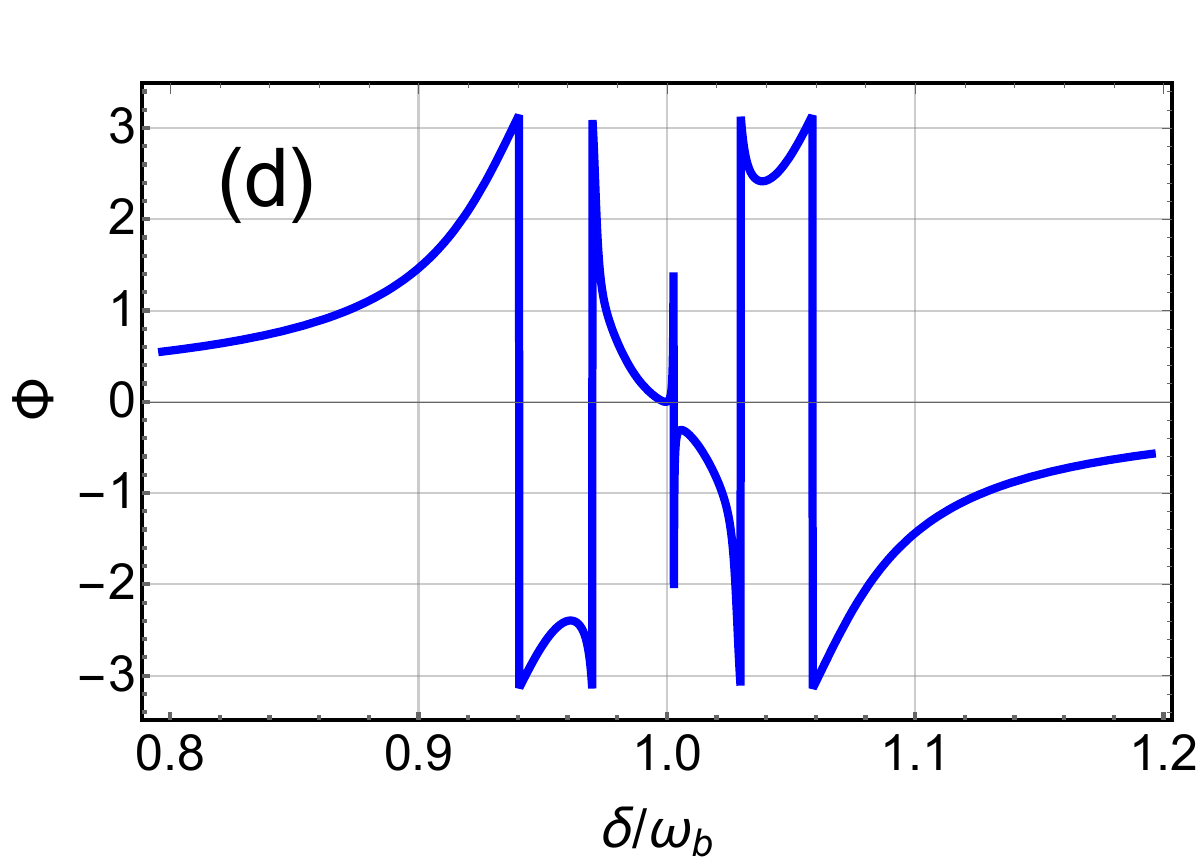}
		\caption{The phase $\Phi$ of the transmitted probe field is plotted against the normalized detuning $ \delta / \omega_b $ for various coupling strengths. (a) $G_{c1}=g_a=0$, $G_{c2}/2\pi=4.8$ MHz, (b) $G_{c1}/2\pi=3.2$ MHz, $G_{c2}/2\pi=4.8$ MHz and $g_a=0$ MHz, (c) $G_{c1}/2\pi=3.2$ MHz, $G_{c2}/2\pi=4.8$ MHz and $g_a/2\pi=2$ MHz, and (d) $G_{c1}/2\pi=3.2$ MHz, $G_{c2}/2\pi=4.8$ MHz and $g_a/2\pi=3$ MHz. In all panels, $\lambda=0.5\kappa_{c}$, $\theta=0$ and remaining parameters are provided in Sect. \ref{I}.} \label{PP}
	\end{center}
\end{figure}

This section investigates the control of a probe light field's transmission to demonstrate both slow and fast light phenomena in an opto-magnomechanical system. Based on Eq. \eqref{Y}, the rescaled transmission field corresponding to the probe field can be expressed as follows
\begin{equation}
	\mathcal{T}=\frac{\epsilon_p-2\kappa_ac_{-}}{\epsilon_p}.
\end{equation}
We define the phase $\Phi$ of the transmitted probe field $\mathcal{T}$ as follows \cite{fi}
\begin{equation}\label{DD}
	\Phi = \operatorname{Arg}[\mathcal{T}].
\end{equation}
Figs. \ref{PP}(a)–\ref{PP}(e) display the phase of the transmitted probe
field $\Phi$ as a function of the normalized detuning $\delta / \omega_b$. In Fig. \ref{PP}(a), both $G_{c1}$ and $g_a$ are turned off, leaving only $G_{c2}$ nonzero, resulting in a typical phase with a single transparency window. When $G_{c1}$ is activated, two transparency windows appear in the phase spectra of the transmitted probe field, as shown in Fig. \ref{PP}(b). With all couplings active, the phase spectrum of the output field is depicted in Figs. \ref{PP}(c) and \ref{PP}(d). These Figures clearly demonstrate that when the microwave-magnon coupling $g_a/2\pi = 2$ $\text{MHz}$, three transparency windows are observed, and when $g_a/2\pi = 2.4$ $\text{MHz}$, four transparency windows are present. In comparison, between Figs. \ref{PP} and \ref{b}, we observe that the phase change is significant and rapid when the absorption peak reaches its maximum in each transparency window.\\
The phase of the transmitted probe field, defined in Eq. \eqref{DD}, plays a crucial role in determining the group delay, which can be expressed as \cite{fi,tau}
\begin{equation} \label{AZ}
	\tau  = \frac{{\partial \Phi }}{{\partial {\omega _p}}} = {\mathop{\rm Im}\nolimits} \left[ \frac{1}{\mathcal{T}}\frac{{\partial \mathcal{T}}}{{\partial {\omega _p}}}\right]. 
\end{equation}
Here, $\tau$ denotes the group delay of the output field. When the group delay is positive $( \tau > 0 )$, the system exhibits slow light behavior, whereas a negative group delay $( \tau < 0 )$ corresponds to a fast light phenomenon.
\begin{figure} [h!] 
	\begin{center}
		\includegraphics[scale=0.4]{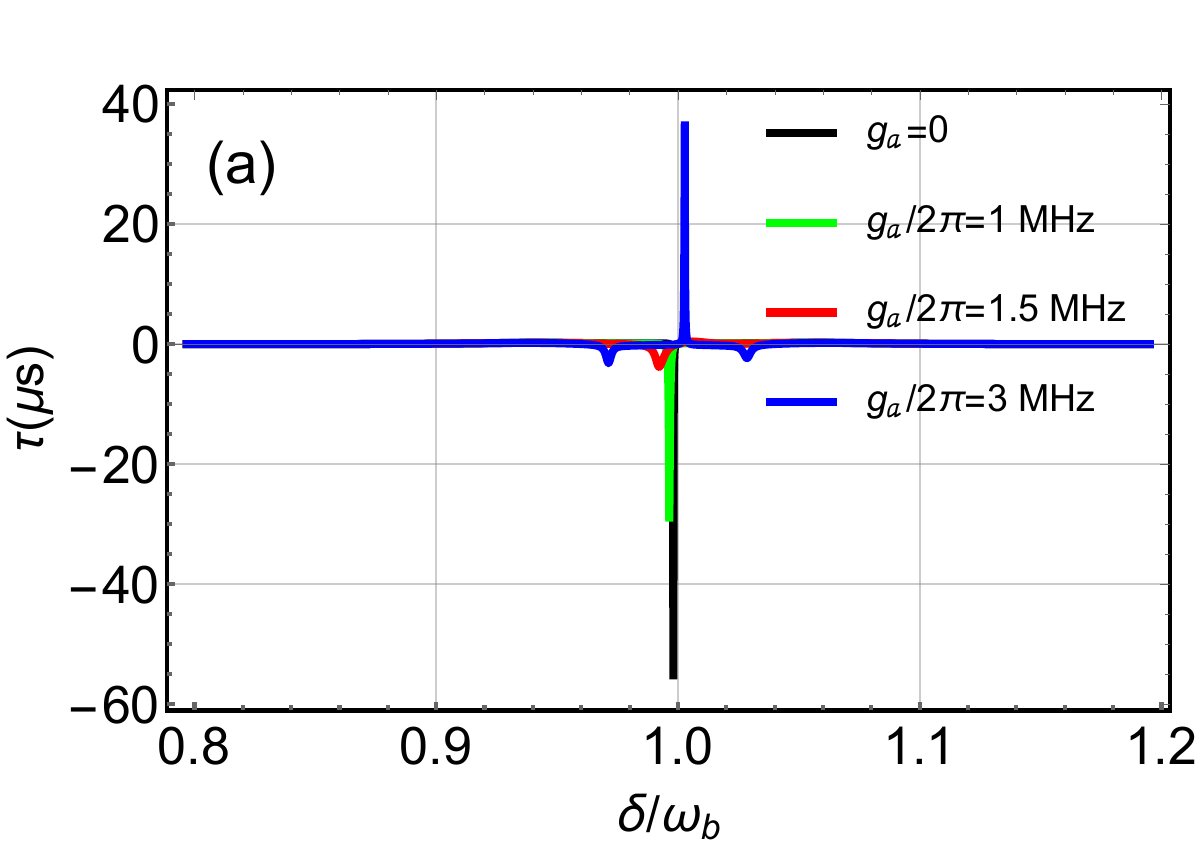}
		\includegraphics[scale=0.4]{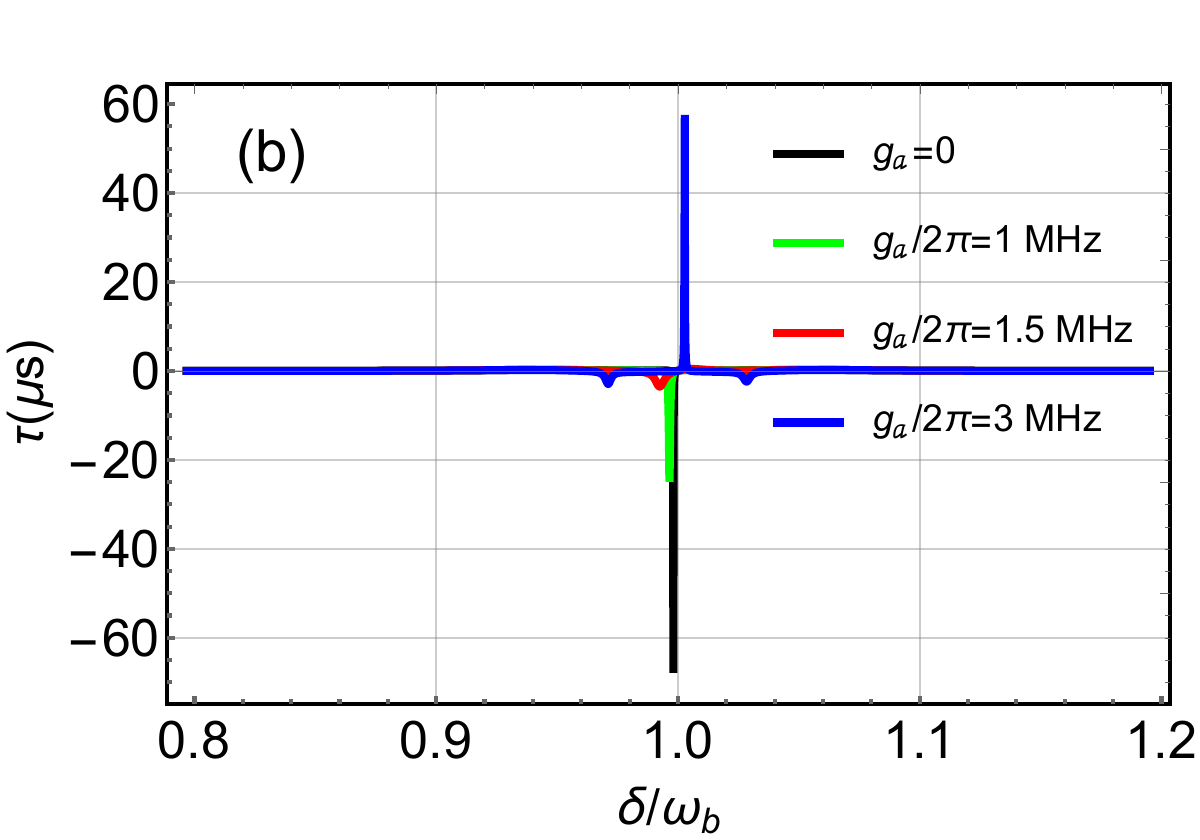}
		\caption{Plot of group delay $ \tau $ versus normalized detuning $ \delta / \omega_b $ for different values of microwave-magnon coupling $ g_a $, with $ {G_c}_1/2\pi=3.5$ MHz, ${G_c}_2 /2\pi={G_m}/2\pi = 4.8 $ MHz and $ \theta= 0 $. (a) $ \lambda = 0 $ and (b) $ \lambda = 0.5\kappa_c $. The remaining parameters are specified in Sect. \ref{I}.} \label{SS}
	\end{center}
\end{figure}
In Fig. \ref{SS}, we present the group delay $\tau$ of the output field at the probe field frequency as a function of the normalized detuning $\delta/\omega_b$, for various values of microwave-magnon coupling $g_a$, while keeping all other parameters constant. When the squeezing parameter is deactivated $(\lambda=0)$ and the microwave-magnon coupling is neglected $(g_a=0)$, we observe only a pronounced fast-light phenomenon in the output probe field. Specifically, at a normalized detuning of $ \delta / \omega_b \approx 0.998 $, the group delay of the output field achieves a negative value of $ \tau \approx -55.79 $ $\mu\text{s}$. When we activated the microwave-magnon coupling at $g_a/2\pi=1$ MHz and $g_a/2\pi=1.5$ MHz, we always notice fast light phenomena, along with a reduction in the negative group delay. For example, at $g_a/2\pi = 1$ MHz, the group delay is  $ \tau \approx -29.64 $ $\mu\text{s}$, while at $g_a/2\pi = 1.5$ MHz, the group delay is  $ \tau \approx -4.42 $ $\mu\text{s}$. At $g_a/2\pi=3$ MHz, we observe that a positive group delay of the output field is achieved at $ \delta / \omega_b \approx 1 $, which corresponds to the slow light effect of the output probe field. The capability to manipulate group delay results from the interaction between the probe field and the anti-Stokes field via quantum interference. This means that we can dynamically adjust the group delay, enabling controllable slow and fast light effects, by varying the coupling strength $g_a$ between microwave and magnons. In addition, the comparison of Figs. \ref{SS}(a) and \ref{SS}(b) shows that increasing the degree of squeezing results in a higher group delay for different coupling strength values $g_a$. Specifically, at $g_a/2\pi = 3$ $\text{MHz}$, the group delay is $ \tau = 37.10 $ $\mu\text{s}$ when $ \lambda = 0 $, and it increases to $ \tau = 57.50 $ $\mu\text{s}$ when $\lambda = 0.5\kappa_c$. By comparing the phase evolution in Fig. \ref{PP} with Fig. \ref{SS} for $g_a/2\pi=3$ MHz, we observe that the peak of the group delay is positive $(\tau>0)$ at $\delta\simeq\omega_b$. This positive peak can be explained by sudden increases of the phase $\Phi$ at $\delta\simeq\omega_b$ (see Eq. \eqref{AZ}), which is illustrated in Fig. \ref{PP}(d).\\
\begin{figure} [h!] 
	\begin{center}
		\includegraphics[scale=0.3]{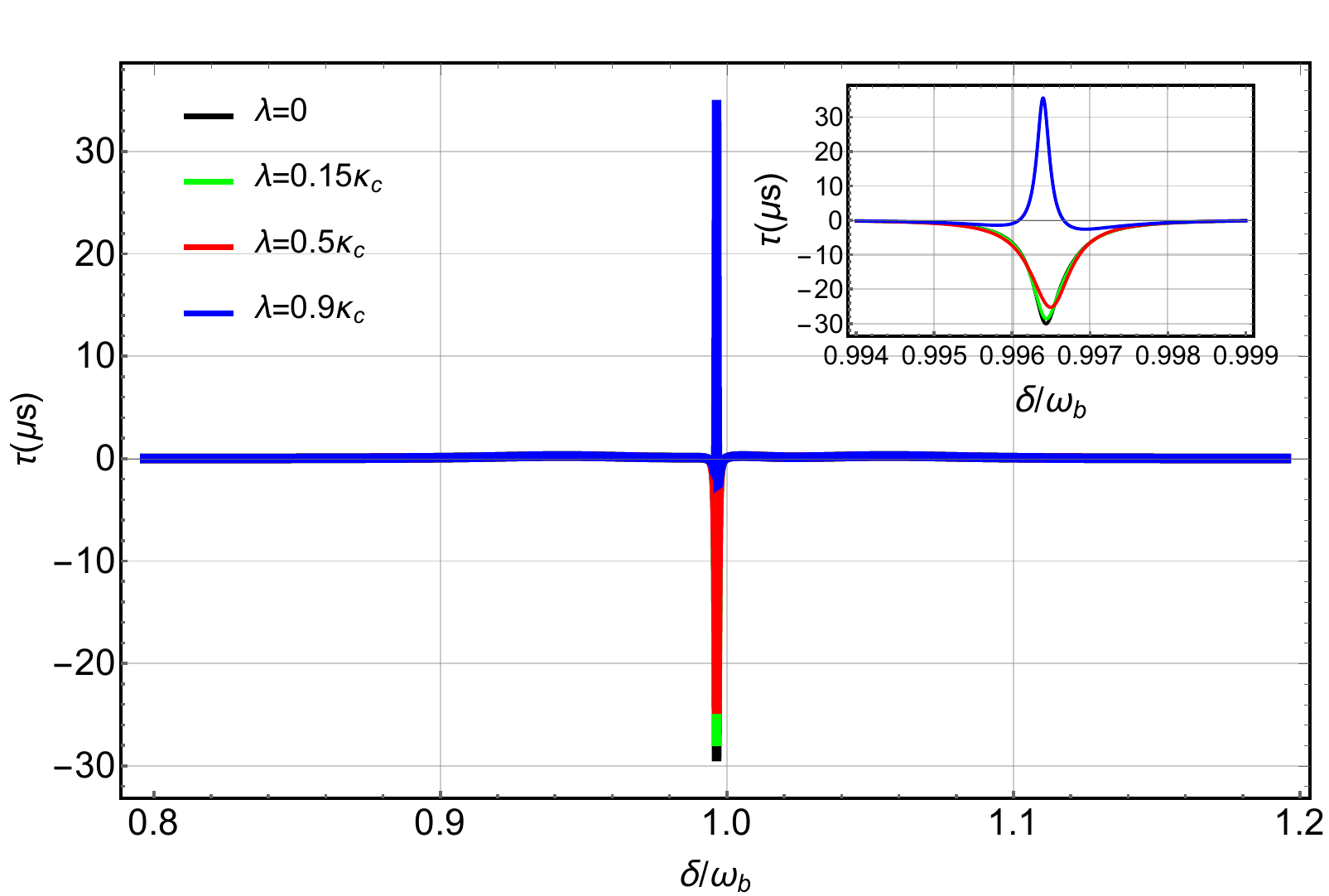}
		\caption{Group delay $ \tau $ versus normalized detuning $ \delta / \omega_b $ for several values of the squeezing parameter $ \lambda $, with $ {G_c}_1/2\pi=3.5$ MHz, ${G_c}_2 /2\pi={G_m}/2\pi = 4.8 $ MHz and $\theta =0 $. The remaining parameters are specified in Sect. \ref{I}.} \label{85}
	\end{center}
\end{figure}
In Fig. \ref{85}, we plot the group delay as a function of the normalized detuning $\delta/\omega_b$ for different values of the squeezing parameter $\lambda$. When the squeezing parameter is set to zero (represented in black), we observe a negative group delay of $\tau=-29.54$ $\mu \text{s}$ at a normalized detuning of $\delta/\omega_b=0.99$, indicating the presence of a fast light effect within the probe field. When we increase the degree of squeezing $\lambda$ to $0.15\kappa_c$ and $0.5\kappa_{c}$, we observe a suppression of the negative group delay experienced by the output field. In blue color, when $\lambda=0.9\kappa_{c}$, we note that the group delay becomes positive, corresponding to the phenomenon of slow light. We note that when we squeeze the magnon in our system, the quantum interference effect between the anti-Stokes field and the probe field generated by the magnon is directly related to the normalized detuning $\delta/\omega_b$ through the squeezing parameter $\lambda$.

\section{CONCLUSION}

In summary, we have studied MMIT, Fano resonance, and slow-fast light effects within an opto-magnomechanical system. We investigated the absorption and dispersion of a weak probe field under the influence of a strong control field. Photon-phonon interactions give rise to OMIT, while phonon-magnon interactions lead to MMIT, and microwave-magnon interactions result in MIT. Our results revealed the existence of four transparency windows in the output probe spectra, arising from interactions between photons, phonons, microwaves, and magnons. We examined how the photon-phonon coupling and the degree of squeezing affect these MMIT windows. Additionally, we thoroughly discussed Fano resonance, which originates from anti-Stokes processes in the system. Furthermore, we explored the conditions for slow and fast light propagation, which can be effectively modulated by tuning key system parameters. Specifically, in our system, it was demonstrated that the group delay can be significantly enhanced by optimizing the microwave-magnon coupling strength and the squeezing parameter.

\appendix
\renewcommand{\thesection}{ \Alph{section}}
\section{Steady-state} \label{B}
The steady-state solutions:
\begin{equation}
	c_s=\frac{\epsilon_L}{\kappa_c+i\tilde{\Delta}_c}, 
\end{equation}
\begin{equation}		 
	a_s=\frac{-ig_am_s}{\kappa_a+i\Delta_a}
\end{equation}
\begin{equation}\label{Am}		 
	m_s=\frac{-ig_aa_s+2\lambda e^{i\theta}m_s^*+\Omega}{\kappa_m+i\tilde{\Delta}_m}, 		
\end{equation}
\begin{equation}
	q_{2s}=\frac{g_2|c_s|^2}{{\omega_b}_2}, 
\end{equation}
\begin{equation}
	q_{1s}=\frac{g_1|c_s|^2-g_m|m_s|^2}{{\omega_b}_1} ,  
\end{equation}
where $\tilde{\Delta}_c=\Delta_c-g_1q_{1s}-g_2q_{2s}$ and $	\tilde{\Delta}_m=\Delta_m+g_mq_{1s}$ are the effective
detuning of the cavity and the magnon, respectively.\\ 
We can write Eq. \eqref{Am} as follows
\begin{equation}\label{An1}
	\begin{aligned}
		m_s=&\frac{2(\kappa_a^2+\Delta_a^2) \lambda \Omega }{A}+\frac{(\kappa_a^2+\Delta_a^2)(\kappa_m-i\tilde{\Delta}_m)\Omega+(\kappa_a+i\Delta_a)\Omega g_a^2}{A},
	\end{aligned}
\end{equation}
with
\begin{equation*}
\begin{aligned}
	A=&\left[(\kappa_m+i\tilde{\Delta}_m)(\kappa_a+i\Delta_a)+g_a^2\right]\left[(\kappa_m-i\tilde{\Delta}_m)(\kappa_a-i\Delta_a)+g_a^2\right]\\
	&+4(\kappa_a^2+\Delta_a^2)\lambda^2,
\end{aligned}
\end{equation*}
If we apply the condition that $|\tilde{\Delta}_{m}|,|\tilde{\Delta}_{c}|,\left|\Delta_a\right| \gg \kappa_m, \kappa_{c},\kappa_{a}$, Eq. \eqref{An1} can be written as:
\begin{equation}
	\begin{aligned}
		m_s=&\frac{2\Delta_a^2 \lambda \Omega }{\left[-\tilde{\Delta}_m\Delta_a+g_a^2\right]\left[-\tilde{\Delta}_m\Delta_a+g_a^2\right]+4\Delta_a^2\lambda^2}\\
		&+i\frac{-\Delta_a^2\tilde{\Delta}_m\Omega+\Delta_a\Omega g_a^2}{\left[-\tilde{\Delta}_m\Delta_a+g_a^2\right]\left[-\tilde{\Delta}_m\Delta_a+g_a^2\right]+4\Delta_a^2\lambda^2},
	\end{aligned}
\end{equation}
In the system under consideration
\begin{equation}
 	\begin{aligned}
 		m_s\approx -i\frac{\Delta_a(\Delta_a\tilde{\Delta}_m - g_a^2)\Omega}{\left[-\tilde{\Delta}_m\Delta_a+g_a^2\right]\left[-\tilde{\Delta}_m\Delta_a+g_a^2\right]+4\Delta_a^2\lambda^2}.
 	\end{aligned}
 \end{equation}
\section{Derivation of $c_-$} \label{A}
$$\alpha_1  =\kappa_c+i(\tilde{\Delta}_c-\delta), \quad \alpha_2=\kappa_c-i(\tilde{\Delta}_c+\delta), \quad \alpha_3=\kappa_a+i(\Delta_a-\delta),$$
$$\alpha_4=\kappa_a-i(\Delta_a+\delta), \quad \alpha_5= \kappa_m+i(\tilde{\Delta}_m-\delta),\quad \alpha_6=\kappa_m-i(\tilde{\Delta}_m+\delta),$$
$$\alpha_7={\omega_b}_1-\frac{\delta}{{\omega_b}_1}(\delta+i{\gamma_b}_1),\quad
\alpha_8={\omega_b}_2-\frac{\delta}{{\omega_b}_2}(\delta+i{\gamma_b}_2),\quad
\mathcal{A}=1+\frac{g_a^2}{\alpha_4\alpha_6}, $$ $$\mathcal{B}=\frac{1}{\alpha_5}+\frac{2\lambda e^{i\theta}}{A.\alpha_5\alpha_6}, \quad
\mathcal{C}=1+\frac{g_a^2}{\alpha_3\alpha_5}-\frac{4\lambda^2}{A.\alpha_5\alpha_6},$$
$$\mathcal{D}=i\alpha_7-\frac{G_{1}^2}{\alpha_2}-\frac{G_{mm}^2}{A.\alpha_6}+\frac{G_{mm}^2B}{C}\left(1-\frac{2\lambda e^{-i\theta}}{A.\alpha_6}\right),$$ $$\mathcal{E}=i\alpha_8-\frac{G_2^2}{\alpha_2},\quad
\mathcal{M}=\frac{1}{D}+\frac{G_2^2}{D.E.\alpha_2}, \quad \mathcal{N}=1-\frac{G_1^2G_2^2}{D.E.{\alpha_2}^2},$$
where $G_1={G_c}_1/\sqrt{2}$, $G_2={G_c}_2/\sqrt{2}$ and $G_{mm}={G_m}/\sqrt{2}$ with ${G_c}_1=i\sqrt{2}g_1c_s$, ${G_c}_2=i\sqrt{2}g_2c_s$ and ${G_m}=i\sqrt{2}g_mm_s$ are the effective magno- and optomechanical coupling strengths, where $|\tilde{\Delta}_{m}|,|\tilde{\Delta}_{c}|,\left|\Delta_a\right| \gg \kappa_m, \kappa_{c},\kappa_{a}$.

\section*{Acknowledgments}

M. Amghar expresses gratitude for the financial support he receives from the National Center for Scientific and Technical Research (CNRST) under the “PhD-ASsociate Scholarship-PASS” program.

\end{document}